\documentclass[a4paper,10pt]{article}
\usepackage[utf8]{inputenc}
\usepackage{graphicx}
\usepackage{color}
\def\vb{{\bf v}}

\def\pa{\partial}
\def\vf{{\bf v}}
\def\Bf{{\bf B}}

\title{Starspots, Stellar Cycles and Stellar Flares: Lessons from Solar Dynamo Models}
\author{Arnab Rai Choudhuri\\
\em Department of Physics\\
\em Indian Institute of Science\\
\em Bangalore -- 560012. India}

\begin{document}

\maketitle

\begin{abstract}

In this review, we discuss whether the present solar dynamo models can be extrapolated to
explain various aspects of stellar activity.  We begin with a summary of the following kinds
of data for solar-like stars: (i) data pertaining to stellar cycles from Ca H/K emission
over many years; (ii) X-ray data indicating hot coronal activity; (iii) starspot data (especially
about giant polar spots); and (iv) data pertaining to stellar superflares. Then we describe
the current status of solar dynamo modelling---giving an introduction to the flux transport
dynamo model, the currently favoured model for the solar
cycle.  While an extrapolation of this model to solar-like stars can explain some
aspects of observational data, some other aspects of the data still remain to be theoretically
explained. It is not clear right now whether we need a different kind of dynamo mechanism
for stars having giant starspots or producing very strong superflares.

\end{abstract}

\section{Introduction}

In elementary textbooks on astrophysics [1, 2], a star is usually modelled
as a non-rotating, non-magnetic spherically symmetric object.  It is the presence of rotation (especially
differential rotation) and magnetic field that makes a 
real star a much more intriguing object, leading to many related
phenomena which we collectively call `stellar activity'.
Although the study of stellar activity has become a thriving research field only within the last
few decades, there is a long history of astronomers studying such activity of the Sun. It was
Galileo [3] who first discovered the solar rotation in 1613 from the changing positions of sunspots
on the solar disk.  When Hale discovered in 1908 [4] that sunspots are regions of strong magnetic
fields, it was the first discovery of magnetic fields in an astronomical system and ushered in a
new era in astronomy with the realization that magnetic fields are ubiquitous in the astronomical
universe. We now know that there is an intimate relation between the rotation and the magnetic field
of a star.  Rotation plays a key role in the dynamo process which presumably generates the magnetic
field.

Since we can resolve the solar surface and observe the magnetic activities taking place there in
considerable detail, solar astronomers have collected data for solar activity for more than a
century. Large sunspots can have sizes of the order of 10,000 km with magnetic field typically
of strength 3000 G.  Well before Hale's discovery of magnetic fields in sunspots, Schwabe 
had noted in 1844 [5] that the number of sunspots seen on the solar surface waxes and wanes periodically.
The sunspot cycle is approximately 11 years and was recognized as the magnetic activity cycle of
the Sun after Hale's discovery of magnetic fields in sunspots.  A major development in solar
astronomy was the realization in the 1940s that the solar corona is much hotter than the solar surface
[6].  The hottest regions of the corona, where temperatures can be more than $2 \times
10^6$ K, usually are found to overlie sunspot complexes, indicating that magnetic fields play
a crucial role in the heating of the corona [7]. Another dramatic manifestation of solar activity
is the solar flare.  First discovered by Carrington in 1859 [8], a large flare can release energy
of the order of $10^{32}$ erg. The fact that flares occur in regions of complex magnetic fields
around sunspots or decayed active regions clearly shows that a 
flare is also caused by the magnetic field and is another
manifestation of the solar magnetic activity.

One intriguing question is whether other stars also have spots, activity cycles, coronae and
flares.  Since a normal star appears as an unresolved point of light even through the largest telescope,
this question cannot be answered by direct observations. However, using ingenious techniques,
astronomers have succeeded in gathering a large amount of information about stellar activity
within the last few decades.  It is found that some stars are much more magnetically active than
the Sun. We have evidence for starspots much larger than the largest sunspots and stellar flares 
much more energetic than the most energetic solar flares.

Along with the observational study of solar activity, considerable amount of theoretical research
has been carried out to understand the different manifestations of solar activity.  Within the
last few years, solar dynamo models have become sufficiently sophisticated and are used now to explain
different aspects of solar activity.  The main question we would like to discuss in the present
review is whether the solar dynamo models can be
extrapolated to other stars and explain their activities.  As
we shall see, some aspects of stellar activity can be explained readily by extrapolating solar
dynamo models.  However, it is not easy to explain very large starspots or very energetic flares
by a straightforward extrapolation of the physics of the Sun.  This raises the question whether
dynamo action in some of these stars is of a qualitatively different nature from the solar
dynamo.  We still do not have a good answer to this question.

Let us begin with a disclaimer.  The author of this review is not an expert on stellar activity
and has only a limited knowledge of this subject.  Still he has undertaken to write this
review because he is not aware of any comprehensive review covering this increasingly
important field of extrapolating solar dynamo models to explain different aspects of stellar
activity. Some early monographs [9, 10] had limited discussions on this subject and
a review by Brun et al.\ [11] covered some aspects. 
Presumably even an incomplete and unsatisfactory survey of this important subject
will be of help to many astronomers, until somebody more 
competent writes a better review.  Within the last few years, there have been some conferences
with the aim of bringing together the two communities working on solar activity and stellar
activity.  On the basis of his experience of attending a few such conferences (notably, IAU
Symposium 273 --- {\em Physics of Sun and Star Spots} [12]; IAU Symposium 286 --- {\em Comparative Magnetic
Minima: Characterizing Quiet Times in the Sun and Stars} [13]), it became clear to the author that
often there is a large communication gap between these communities.  Hopefully a solar physicist
not knowing much about stellar activity will get an idea from this review of solar-like activity
phenomena in other stars.  On the other hand, stellar astronomers not knowing much about the
recent advances in solar dynamo theory will form an idea where the theoretical efforts stand
now. 

The next Section summarizes the salient features of observational data related to stellar
activity.  Then \S~3 gives an introduction to the flux transport dynamo model, the currently
favoured theoretical model of the solar cycle.  Afterwards we shall discuss in \S~4 whether the flux
transport dynamo model can be extrapolated to solar-like stars to model their activity.  Our
conclusions are summarized in \S~5. 

\section{Observational data of stellar activity}

The first indication that some stars are magnetically active came from observations in Calcium
H/K lines. These lines form in the chromosphere somewhat above the stellar photosphere where 
the optical depth for these lines becomes
$\approx 1$.  If this region has a temperature less than the photosphere, then we expect
absorption lines in Ca H/K.  However, if this region gets heated up due to the presence of
the magnetic activity, then there can be an emission core. Research in the field of stellar 
activity began with the discovery by Eberhard and Schwarzschild in 1913 [14] that the spectra of
some stars show emission in Ca H/K. It was later found by Wilson and Bappu [15] that the
width of the Ca H/K lines is correlated with the absolute magnitude of the star---this correlation
being now known as the {\em Wilson--Bappu effect}. Stellar chromospheric activity has been
reviewed by Hall [16].

Before we start discussing the observational data in more detail, we would like to
point out one important physical effect.  The dynamo process which generates the magnetic
activity in stars is powered by convection taking place inside the stars.  So we expect
the magnetic activity to be visible at the surface in those stars which have an outer
shell of convection below their surfaces.  This is the case for the late-type stars occupying
the right side of the Hertzsprung--Russell (HR) diagram. On the other hand, the early-type
stars occupying the left side of the HR diagram have convective cores.  Even if the dynamo
process takes place in the core, the magnetic field is unlikely to come out through the
stable surrounding layers having high electrical conductivity. That is why we expect to
find evidence for stellar activity primarily in the late-type stars.  We shall see that
this expectation is borne out by observational data.

\begin{figure}
%\vspace{8cm}
\centerline{\includegraphics[width=0.6\textwidth,clip=]{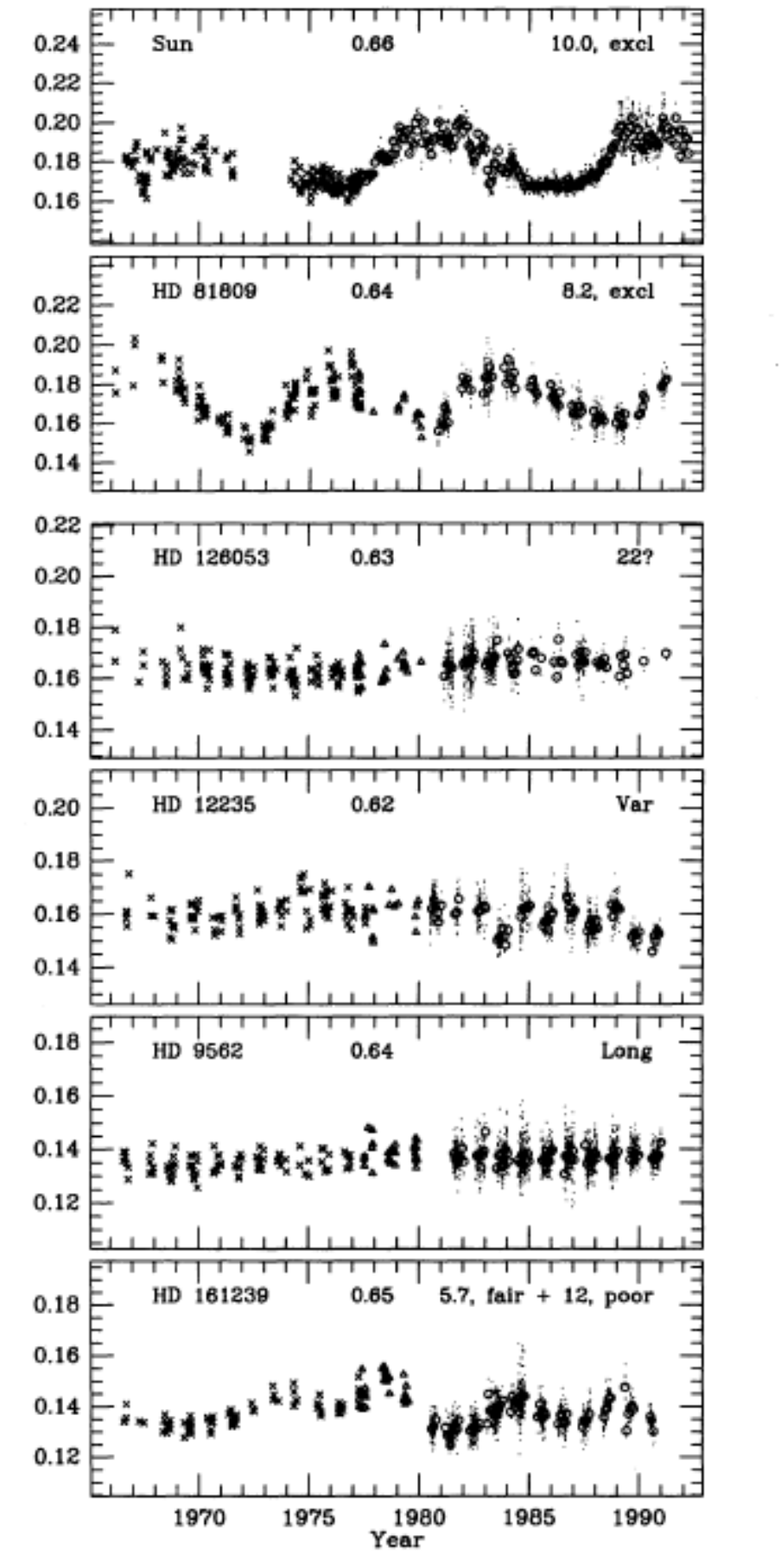}}
\caption{The variation of Ca H/K emission with time for the Sun and a 
few stars.  From Baliunas et al.\ [18].}
\end{figure}

What we present below is a very incomplete survey of a few selective aspects of stellar
activity to which dynamo modellers should pay attention.  The subject of asteroseismology,
which may play an important role in stellar dynamo modelling in future, is outside the
scope of this review (see [11] for the current status of asteroseismology). We do not make
any attempt of providing a comprehensive bibliography.  Rather, the discussion is centred
around a few key papers of historical importance.

\subsection{Stellar activity cycles}

If other solar-like stars also have activity cycles like the 11-year solar cycle, then we
would expect the Ca H/K emission to vary with the stellar cycle.  We need to monitor the
Ca H/K emission from a star for several years in order to find out if the star has a cycle.
In the 1960s Olin Wilson (of Wilson--Bappu effect fame) started 
an ambitious programme at Mount Wilson Observatory 
of monitoring Ca H/K emission from a large number of stars.  After collecting data for
several years, Wilson reported the discovery of stellar cycles [17].
The most comprehensive presentation of data from this project can be found in the paper
by Baliunas et al.\ [18] published shortly after Wilson's death. Figure~1 is a panel from
this paper showing the variation in Ca H/K emission with time for several stars. 
Many stars were found to have regular periods.  Some stars showed more irregular variations
in Ca H/K indicating the existence of grand minima as in the Sun (such as the Maunder
minimum in the seventeenth century). 

\begin{figure}
%\vspace{8cm}
\centerline{\includegraphics[width=0.75\textwidth,clip=]{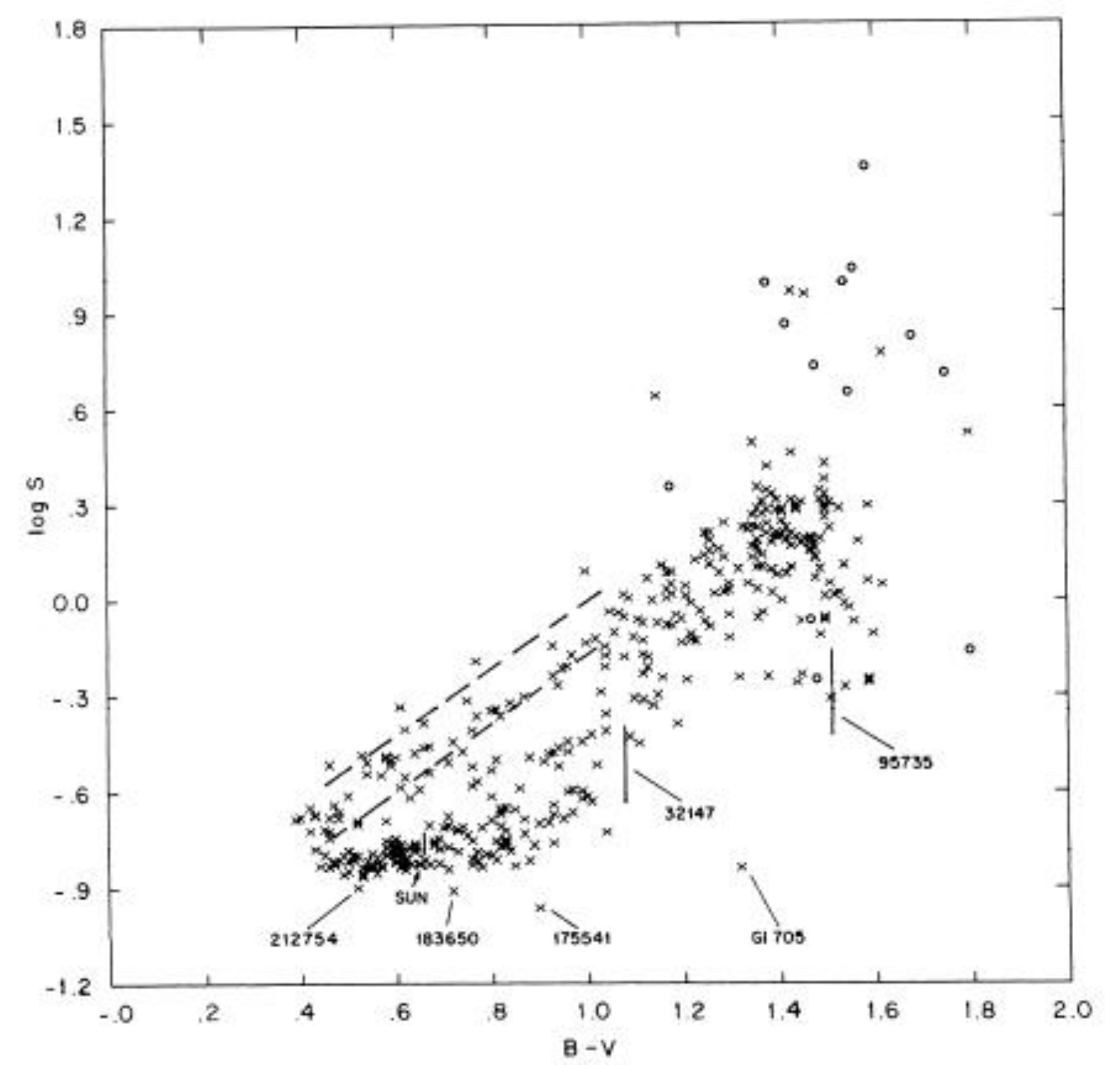}}
\caption{The averaged
Ca H/K emission from many stars plotted against their spectral type. From Vaughan and Preston [19].}
\end{figure}

The Ca H/K emission averaged over a few years is a good index of a star's magnetic
activity.  Figure~2 taken from a paper by Vaughan and Preston [19] shows the averaged
Ca H/K emission from many stars plotted against their spectral type. As we already
pointed out, we expect visible manifestations of magnetic activity mainly from the
late-type stars and this is what is seen in Figure~2. Very curiously, one sees a gap in
Figure~2 between two bands of data points.  This is known as the {\em Vaughan--Preston
gap} and the theoretical reason behind it is still not fully understood.

\begin{figure}
%\vspace{8cm}
\centerline{\includegraphics[width=0.7\textwidth,clip=]{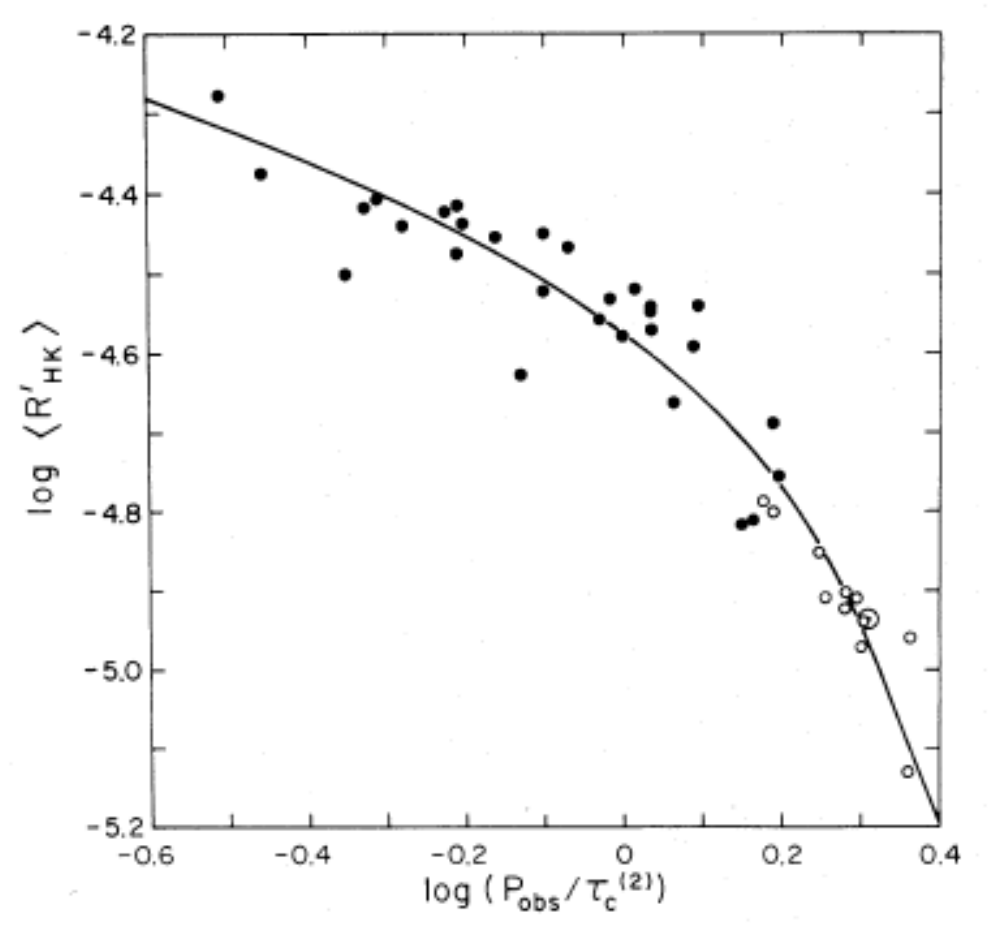}}
\caption{The averaged
Ca H/K emission from many stars plotted against the Rossby 
number, defined as the ratio of the rotation period and the convective turnover time. From Noyes et al.\ [20].}
\end{figure}

If the dynamo process producing the magnetic activity depends on rotation, then we
would expect more rapidly rotating stars to have stronger magnetic activity. This was
established by Noyes et al.\ in 1984 [20]. When they plotted the averaged Ca H/K data against
rotation period, they found that there was statistically more Ca H/K emission from stars
with shorter rotation period, but there was considerable scatter in the plot.  Very
intriguingly, when they plotted the averaged Ca H/K emission against the Rossby 
number (i.e.\ the ratio of the rotation period and the convective turnover time),
the scatter was significantly reduced.  Figure~3 reproduces a famous plot from their
paper. There is evidence that stellar rotations slow down with age [21]---presumably
as a result of stars losing angular momentum through stellar wind, like what is believed
to happen in the Sun. Though there is likely to be a spread in rotation periods when
new stars are born, a longer rotation period (the right side of Figure~3) 
statistically implies an older star.
So Figure~3 can be viewed as a plot showing that stellar activity decreases with stellar
age.

\begin{figure}
%\vspace{8cm}
\centerline{\includegraphics[width=0.6\textwidth,clip=]{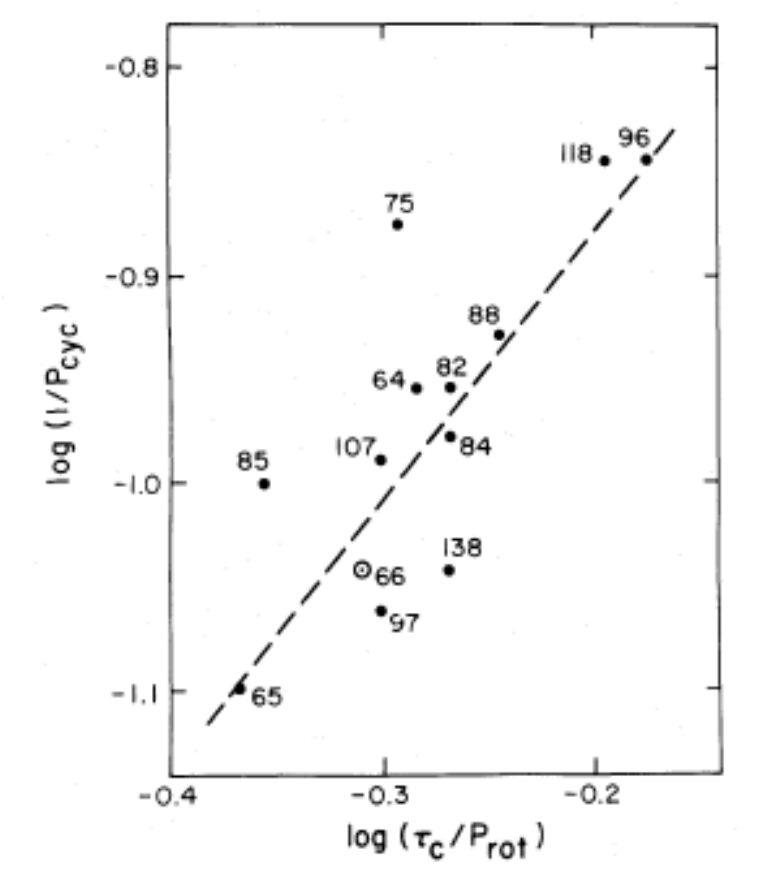}}
\caption{The plot of inverse cycle period against the inverse of Rossby number for
several stars. From Noyes, Weiss and Vaughan [22].}
\end{figure}

If the Ca H/K emission from a star is roughly periodic, one can measure the stellar
activity period and study its relation with the stellar rotation period
[22, 23].  Figure~4 taken from Noyes, Weiss and Vaughan [22] is a plot of
inverse cycle period against the inverse of Rossby number.  It is found that stars with
shorter rotation periods tend to have shorter activity cycle periods.  As we shall see later,
this observational result has proved particularly challenging to explain with the
flux transport dynamo model, although it could be explained very easily with the
older $\alpha \Omega$ dynamo model [22].

\subsection{Stellar coronae and X-ray emission}

A hot gas having temperature of the order of a million degrees is expected to emit
X-rays. The first X-ray images of the hot solar corona were obtained by Skylab in
the early 1970s. More recent space missions like Yohkoh, SoHO, TRACE and Hinode have
provided spectacular images of the solar corona obtained in X-rays or extreme
ultraviolet. For other stars, we cannot expect to image their coronae.  But, if
they have hot coronae, then we can expect to detect X-ray emission from the stars.

\begin{figure}
%\vspace{8cm}
\centerline{\includegraphics[width=0.6\textwidth,clip=]{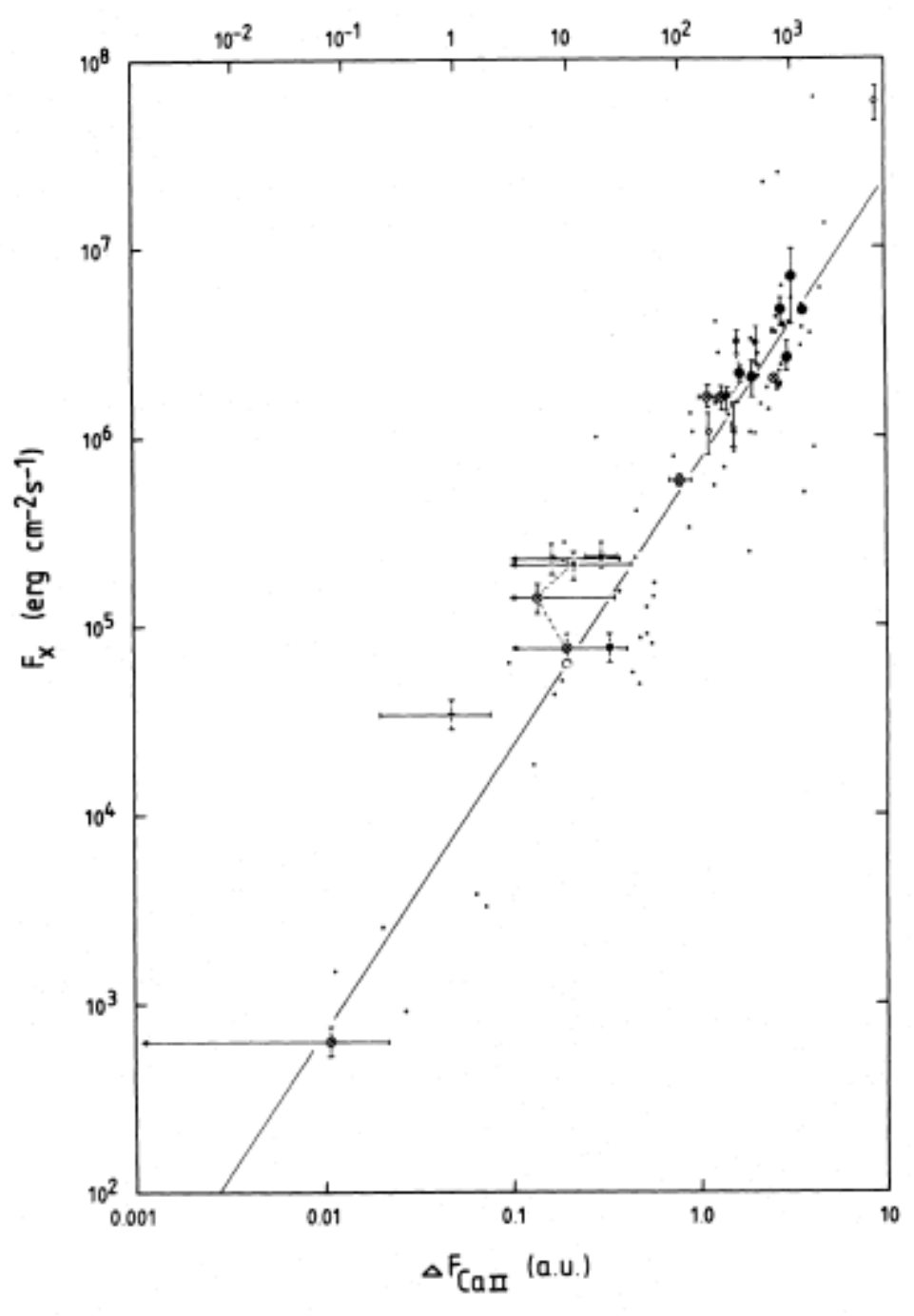}}
\caption{The correlation between the X-ray luminosity and the time-averaged
emission in Ca H/K for several stars. From Schrijver, Dobson and Radick [25].}
\end{figure}

The Einstein X-ray Observatory was able to detect X-ray emission from many
stars: both early-type and late-type [24]. For the early-type stars, the strength of
X-ray emission was found to be proportional to luminosity, suggesting that the
X-ray emission from these stars was related to their overall structure and not
to their magnetic activity.  In the case of late-type stars, however, their X-ray
brightness was found to be correlated with the Ca H/K emission [25], as seen in Figure~5.
This clearly indicates that the X-ray emission in these stars is related to
magnetic activity like the Ca H/K emission and presumably must be coming from
the hot coronae.

\begin{figure}
%\vspace{8cm}
\centerline{\includegraphics[width=0.7\textwidth,clip=]{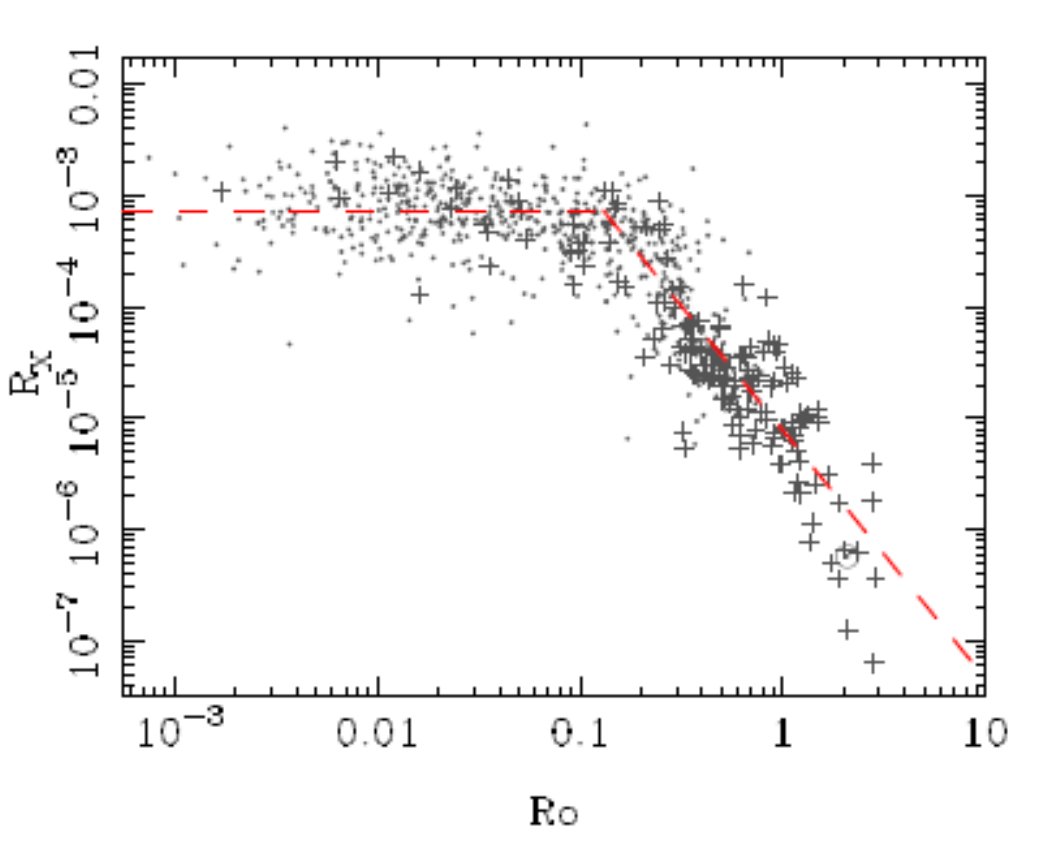}}
\caption{The plot of X-ray luminosity against the Rossby number for several
late-type stars.  From Wright et al.\ [27].}
\end{figure}

Just as stars with low Rossby number (i.e.\ more rapidly rotating) have higher
Ca H/K emission, they are expected to have more extensive coronae and stronger
X-ray emission as well.  This was actually found [26, 27] as shown in Figure~6, where X-ray brightnesses of
different late-type stars are plotted against their Rossby number. We find the
data points distributed around a curve with only a modest scatter, as in the case
of Figure~3, which was a similar plot with Ca H/K emission rather than X-ray emission.

\subsection{Starspot imaging}

Can we map spots on the surface of a star which we are unable to resolve?
The very ingenious technique of Doppler imaging, pioneered by Vogt and
Penrod in the early 1980s [28], has now made this possible. Figure~7 explains this technique.
If a star is rotating around its axis, the part moving towards us will have
spectral lines blue-shifted, whereas the part moving away will have them
red-shifted.  Since the star is not resolved by the telescope, the net result
is the broadening of the spectral line.  Now suppose there is a large starspot.
When it is in the part moving towards us, there is less contribution to
the blue-shifted part of the spectral line.  As a result, there will be a
bump in the blue-shifted part of the  spectral line. As the starspot moves
across the surface of the star and goes to the part moving away, the bump
will move towards the red-shifted part of the  spectral line.  By analyzing
the movement of the bump across the spectral line, it is possible to figure out
the size and the location (i.e.\ the latitude) of the starspot.

\begin{figure}[t]
%\vspace{8cm}
\centerline{\includegraphics[width=0.55\textwidth,clip=]{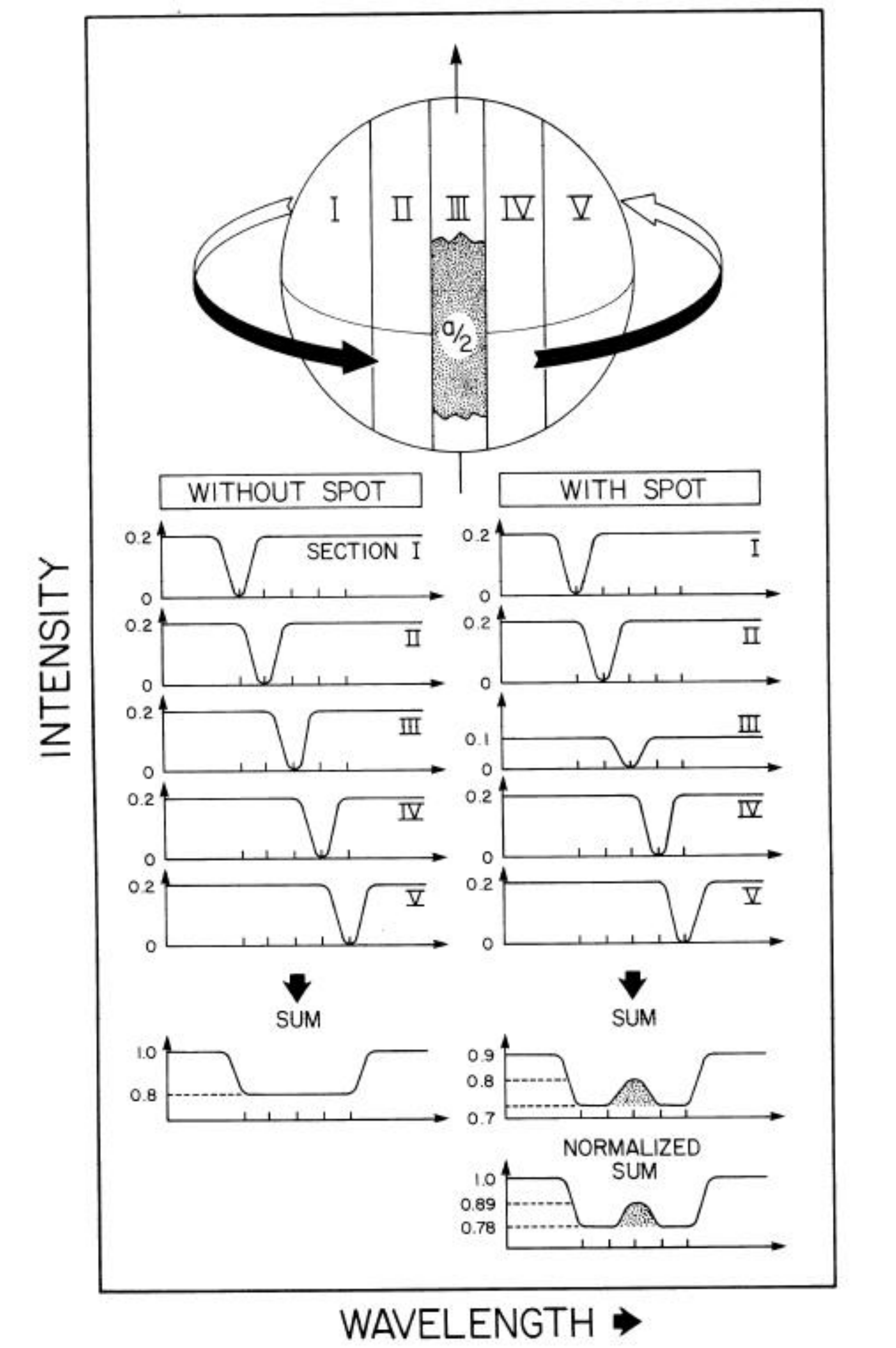}}
\caption{A schematic illustration to explain the Doppler imaging
technique.  The spectra from the five parts of a rotating star are
indicated for the two cases: (i) when there is no spot, and (ii) when
there is a spot in the middle.  The spectrum of the unresolved star is
the sum of the contributions from all these different parts.  From Vogt and Penrod [28]. }
\end{figure}

Through this Doppler imaging technique, many stars have been found to have
very large starspots.  What is more remarkable, these giant starspots are often
found in the polar regions of stars---especially in the case of rapidly rotating
stars. Figure~8 shows a giant polar starspot mapped
by Strassmeier [29]. We shall later discuss the possible reasons behind these
starspots appearing near the poles, in contrast to sunspots which usually appear
within 40$^{\circ}$ of the solar equator. The field of observational study of
starspots has really blossomed in the last few years. The evolution and decay
of giant starspots have been studied [30].  It has also become possible
to do magnetic field measurements of starspots through the Zeeman--Doppler
technique [31]. The readers are referred
to the comprehensive reviews by Berdyugina [32] and Strassmeier [33] for an
account of the field of starspots, whereas the subject of magnetic field
measurement of late-type stars is reviewed by Reiners [34].

\begin{figure}[t]
%\vspace{5cm}
\centerline{\includegraphics[width=0.6\textwidth,clip=]{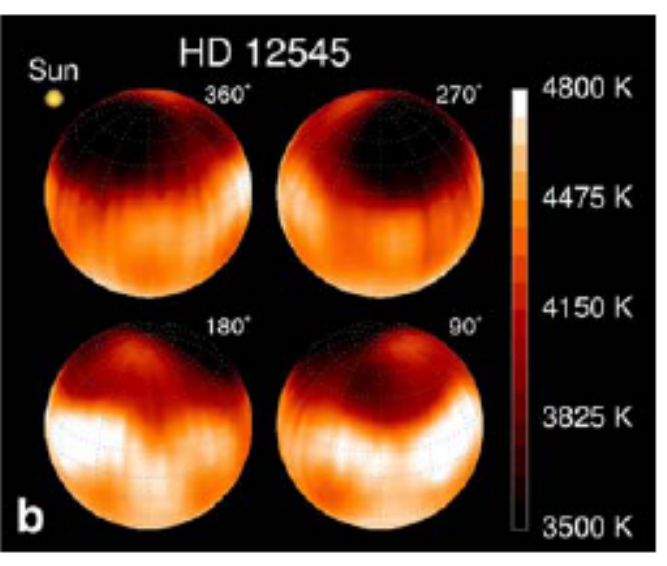}}
\caption{A giant polar starspot mapped by the Doppler imaging technique. This is
the colour version (given in [33]) of the original figure from
Strassmeier [29].}
\end{figure}

Differential rotation plays a crucial role in the dynamo process.  It has been
possible to estimate the surface differential rotation of some stars from the 
study of starspots [35]. Another intriguing aspect of starspots which should be a
challenge for dynamo modellers is the so-called `flip-flop': the observational
indication that some stars have starspots separated by 180$^{\circ}$ in longitude
which alternate in strength [36, 37].

\subsection{Stellar flares}

The most powerful solar flares release energy of the order of at most $10^{32}$
erg. Curiously, the first solar flare recorded by any human being, the Carrington
flare of 1859 [8], has so far remained the most powerful flare
recorded and presumably had an energy of such magnitude. Even at its peak, such
a flare would not increase the overall brightness of the Sun by more than 1\%.
It will be extremely hard to detect such a flare in a distant star.  Only when
the stellar flare is a much more energetic superflare ($\approx$ 10$^{33}$ -- 10$^{34}$ erg)
and leads to an appreciable temporary increase in the luminosity
of a star, we have a good chance of detecting it.  Since there is no way of
knowing in advance when a stellar flare is going to take place, the first detections
of stellar flares were serendipitous detections when a superflare occurred while a star
was being observed [38].

A systematic study of stellar flares became possible only after the launch of
the Kepler mission aimed at discovering exoplanets.  This mission
continuously monitors the brightness of about 145,000 stars in a fixed field of view.
Analyzing the data of this mission, Maehara et al.\ [39] reported the discovery
of 365 superflares. Figure~9 shows the brightness variations of two stars in which
superflares were seen.  Initially it was thought that the occurrence of a superflare
required a close binary companion, like a `hot Jupiter' (i.e.\ a nearby massive
planet).  However, it was found that 14 of these superflares took place in slowly
rotating stars, which do not have any close companions.

\begin{figure}
%\vspace{8cm}
\centerline{\includegraphics[width=0.65\textwidth,clip=]{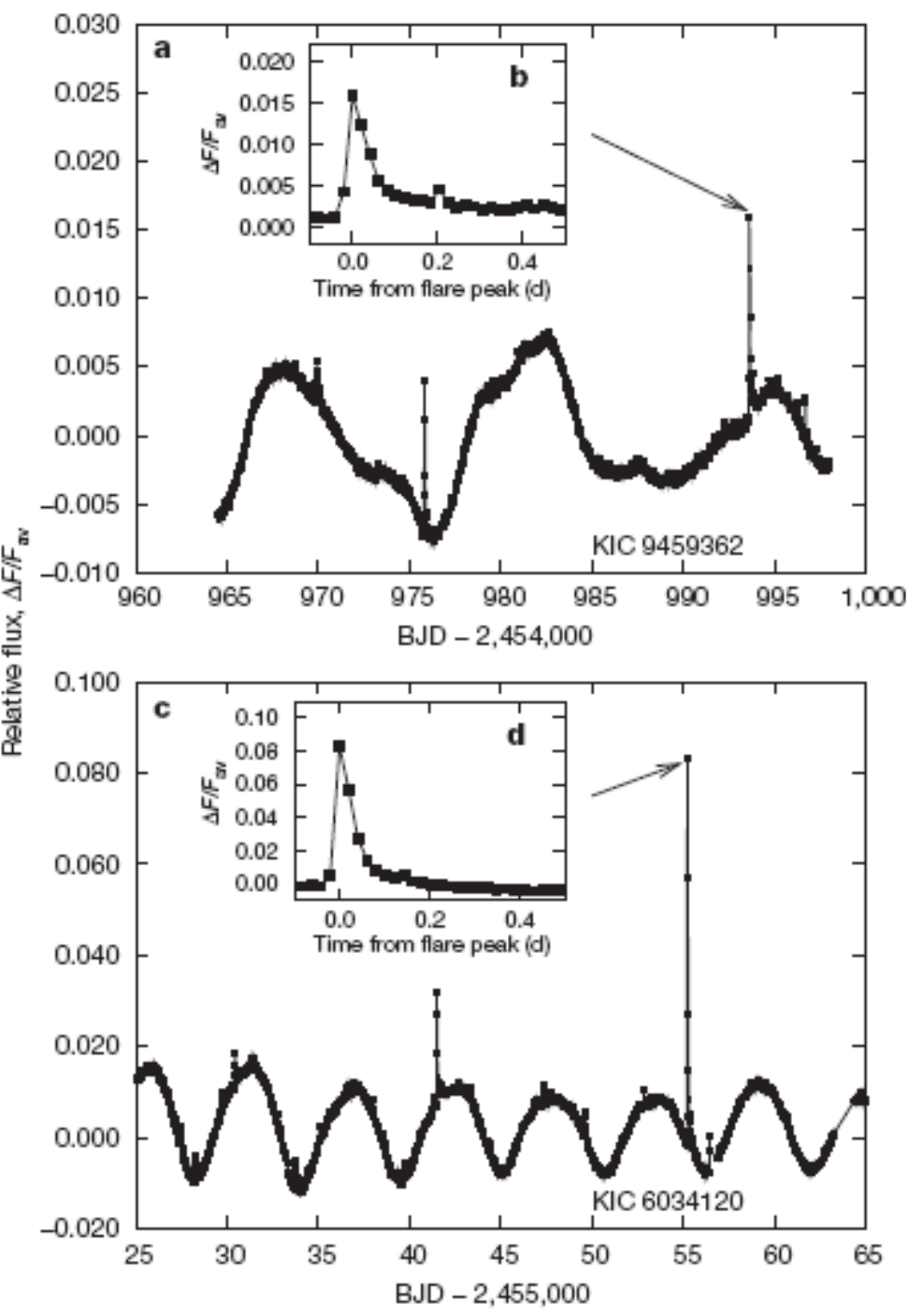}}
\caption{The brightness variations of two stars in which
superflares were seen. From Maehara et al.\ [39].}
\end{figure}

Although the number 365 of stellar superflares is not very large to do a completely
reliable statistical analysis, one can still study the occurrence statistics. Figure~10
taken from Shibata et al.\ [40] shows the occurrence 
statistics of these superflares in the same plot with the occurrence
statistics of solar flares as well as microflares and nanoflares occuring on the
Sun.  All these different kinds of flares seem to obey the same power law
to a good approximation.

\begin{figure}
%\vspace{8cm}
\centerline{\includegraphics[width=0.75\textwidth,clip=]{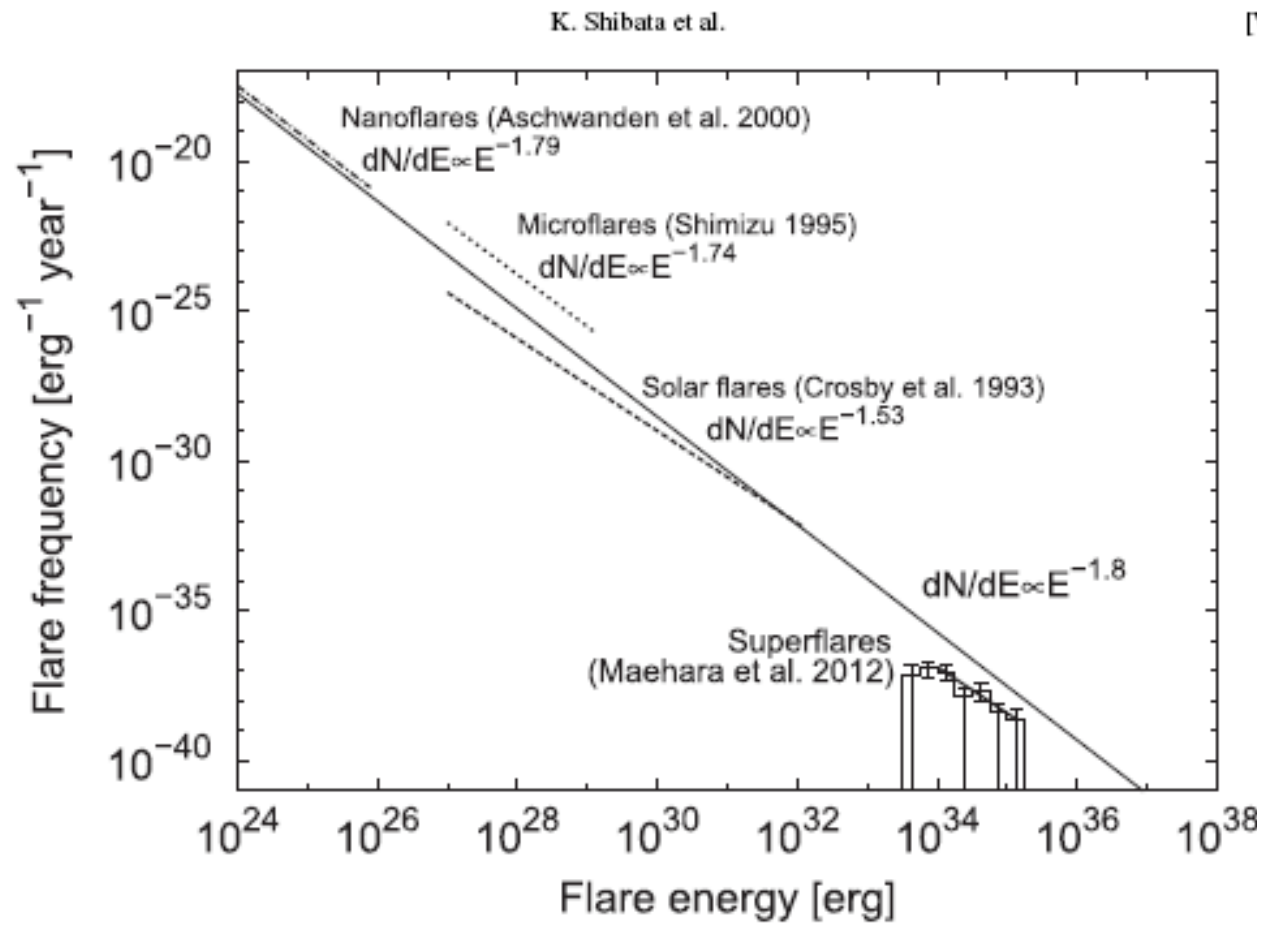}}
\caption{The frequency of various kinds of flares plotted against the
flare energy. From Shibata et al.\ [40].}
\end{figure}

The most powerful solar flare in recent times was the 1989 flare which caused a 8-hour
power blackout in Quebec (the regions around the geomagnetic pole being affected the
maximum).  The 1859 Carrington flare has been estimated to be about three times more
powerful than the 1989 flare [41].  These flares involved energy of order 10$^{32}$ erg.
If one believes the occurrence statistics based on 14 superflares that occurred in
slowly rotating solar-like stars, then a flare of energy 10$^{34}$ erg is expected in
800 years and a flare of energy 10$^{35}$ erg in 5000 years [39].  If
flares of such strength do occur and affect the Earth, it will have a disastrous effect
on our current technology-dominated human civilization.  Hence it is a very important
question whether such superflares can occur on our Sun.  We shall come back to this
question later.

\section{The flux transport dynamo model of the solar cycle}

After summarizing the salient features of stellar magnetic activity, we now introduce
the flux transport dynamo model of the solar cycle, before coming to the question
of providing theoretical explanations for different aspects of stellar activity
in the next Section.
It is the nonlinear interaction between the magnetic field and the velocity field
within the solar convection zone which sustains the solar magnetic field and produces
the solar cycle.  One of the remarkable developments in solar physics within the
last few decades has been helioseismology, which has provided a huge amount of
information about large-scale flows in the solar convection zone such as the differential
rotation and the meridional circulation.  It is the lack of such detailed information about
the flow fields inside stars which hampers the development of stellar dynamo models.

\subsection{The toroidal and the poloidal magnetic fields of the Sun}

Sunspots often appear in pairs at approximately the same solar latitude.  When Hale
et al.\ [42] discovered in 1919 that the two sunspots of a typical pair have opposite magnetic
polarity (the polarity sense being opposite in the two hemispheres), it became
clear that there must be some toroidal magnetic flux system underneath the solar
surface from which magnetic strands rise to the solar surface to produce the bipolar
sunspots.  We, therefore, assume the sunspots to be a proxy for the toroidal
field.  With the discovery of the much weaker magnetic field near the Sun's polar
region by Babcock and Babcock [43], it was established that the Sun has a poloidal
magnetic field as well.  Now we know that the polar field appears weak (about 10 G)
only in low-resolution magnetograms, but is actually concentrated inside magnetic
flux bundles to strength of order 1000 G [44].  
The theoretical understanding of why the magnetic
field at the solar surface appears intermittent comes the study of magnetoconvection,
first pioneered by Chandrasekhar [45].  Further work by Weiss [46] and others
established that the interaction with convection makes magnetic field confined within
flux concentrations.  It is believed that the magnetic field exists in the form of
magnetic flux tubes throughout the solar convection zone.

In a pathbreaking work in 1955, Parker [47] developed the scenario that the toroidal and
the poloidal magnetic fields of the Sun sustain each other.  Although we now believe
that some important modifications of Parker's ideas are needed, the overall scenario
of the toroidal and poloidal fields sustaining each other gets support from the observational
data of the solar polar magnetic fields.  Figure~11 shows the time variation of the magnetic
fields at the two poles of the Sun, along with the sunspot number plotted below. 
It is seen that the sunspot number, which is an indication of the
strength of the toroidal component, becomes maximum around the time when the polar
field (the manifestation of the poloidal component) is close to zero.  On the other
hand, the polar fields are strongest when the sunspot number is close to zero.

\begin{figure}[t]
%\vspace{8cm}
\centerline{\includegraphics[width=0.9\textwidth,clip=]{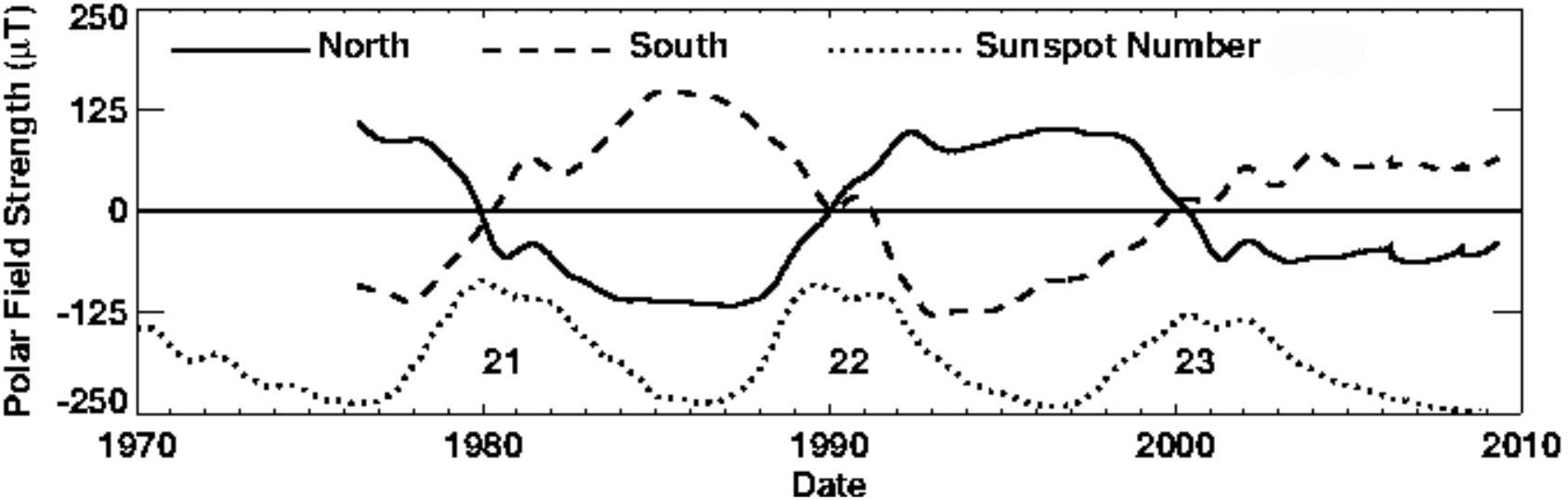}}
\caption{The evolution of the Sun's polar fields with time, with the sunspot
number plotted at the bottom.  From Hathaway [48].}
\end{figure}

\subsection{The generation and the dynamics of the toroidal field}

In order to have the kind of oscillation between the toroidal and poloidal fields
seen in Figure~11, we need processes to generate the toroidal field from the poloidal
field and to generate the poloidal field from the toroidal field.  Since we are
dealing with a high magnetic Reynolds number situation and the magnetic field is
approximately frozen in the plasma (see, for example, Choudhuri [49]), differential
rotation is expected to stretch out the poloidal field to produce the toroidal
field.  Since helioseismology has discovered that the region of strong differential
rotation (known as the tachocline) is concentrated near the bottom of the convection
zone [50], we expect the strong toroidal field to be generated there.  Interaction with
convection presumably keeps this toroidal field concentrated within toroidal magnetic
flux tubes.  Parker [51] proposed the idea of magnetic buoyancy that the magnetic
pressure inside the flux tube may make it expand and cause a decrease in density,
making the flux tube buoyant.  It is found that the magnetic buoyancy is particularly
destabilizing within the convection zone, but is suppressed to a large extent below
its bottom [52, 53]. Since the toroidal flux tube is
created by the differential rotation of the tachocline exactly at the bottom of
the convection zone, a part of it may come within the convection zone, become buoyant
and rise through the convection zone to produce the bipolar sunspots, whereas other
parts may remain anchored slightly below the bottom of the convection zone. In order
to understand how the bipolar sunspots form, one then has to study the dynamics
of the part of the flux tube that has come within the convection zone, which can
be done with the help of the thin flux tube equation [54, 55].

Although the two sunspots in a bipolar pair appear approximately at the same latitude,
a more careful study shows that the leading sunspot is statistically found slightly
closer to the solar equator.  The tilts of the bipolar sunspot pairs are found
to increase with latitude---a result known as Joy's law, after Hale's collaborator
Joy who established this law in their pioneering study of sunspot pairs
[42]. Presumably this tilt is produced by the action of the Coriolis
force on the rising flux tube.  Choudhuri and Gilman [56] and Choudhuri [57]
were the first to study the effect of the Coriolis force on magnetic buoyancy and
found that the Coriolis force plays a much more important role in this problem than
recognized hitherto.  If we assume the magnetic energy to be in equipartition with
the kinetic energy of convection, then the magnetic field inside the flux tube at the
bottom of the convection zone cannot be larger than $10^4$ G. Flux tubes with such
magnetic field strengths are diverted to rise parallel to the rotation axis and
emerge at high latitudes.  As we shall discuss later, this result is likely to
have important implications for polar starspots.  Only if the magnetic field inside
flux tubes is as strong as $10^5$ G, the flux tubes can come out radially, making
the appearance of sunspots at low latitudes possible.  D'Silva and Choudhuri [58]
also found that they could fit Joy's law with their simulation only if the magnetic
field inside the flux tubes at the bottom of the convection zone was of order
$10^5$ G.  Soon confirmed by other authors [59, 60],
this result puts an important constraint on the magnetic field inside the Sun
and imposes a constraint on possible dynamo mechanisms, as we shall see. Although
some effects have been postulated that can suppress the Coriolis
force [61, 62], it is not clear if these effects would be important in the interior
of the Sun and the initial magnetic field inside rising flux tubes presumably has
be of order $10^5$ G in order to match observations.

\subsection{The generation of the poloidal field}

If we start from a poloidal field, we have discussed how the differential rotation
can stretch it out to create the toroidal field in the tachocline and then how parts
of this toroidal field can rise in the form of flux tubes to produce the bipolar sunspots.
In order to have the magnetic cycle encapsulated in Figure~11, we now need a mechanism
for producing the poloidal field from the toroidal field.  The early idea due to
Parker [47] and then elaborated by Steenbeck, Krause and R\"adler [63] was that the turbulence
within the Sun's convection zone would involve helical fluid motions due to the presence
of the Coriolis force arising out of the solar rotation and that this helical turbulence
would twist the toroidal field to produce the poloidal field.  This mechanism, christened
as the $\alpha$-effect, can work only if the toroidal field is not too strong so that
it can be twisted by turbulence. When the flux tube simulations of sunspot formation
described in \S~3.2 suggested that the toroidal field at the bottom of the convection zone
is much stronger than the equipartition value, it became clear that the $\alpha$-effect
could not twist such a strong field.

Another alternative mechanism for the generation of the poloidal field was proposed
by Babcock [64] and Leighton [65].  Since the two opposite-polarity sunspots in a
bipolar pair form at slightly different latitudes, they pointed out that the decay of these
sunspots would cause magnetic flux of opposite polarities to be spread out at slightly
different latitudes, giving rise to a poloidal field.  So we can view a bipolar sunspot
pair as a conduit for converting the toroidal field to the poloidal field.  It forms
due to the buoyant rise of the toroidal field and we get the poloidal field after its
decay. Now, one requirement of solar dynamo models is that we should have something
like a dynamo wave propagating equatorward, in order to explain the  the appearance
of sunspots at increasingly lower latitudes with the progress of the cycle (as
encapsulated by the well-known butterfly diagram). If the toroidal
field is produced by the differential rotation as mapped by helioseismology and the
poloidal field is produced by the Babcock--Leighton mechanism, then it is found that
the dynamo wave would propagate poleward, creating sunspots at higher latitudes with
the progress of the solar cycle.  So we need something else to make the theory fit with
the observations.

We believe that this additional something is provided by the meridional circulation of
the Sun.  It has been known for some time that there is a poleward flow of plasma
at the solar surface having an amplitude of order 20 m s$^{-1}$. Since we do not
expect the plasma to pile up at the poles, there has to be a return flow underneath
the Sun's surface to bring back the plasma to the equatorial region.  This meridional
circulation presumably arises from turbulent stresses within the Sun's convection zone.
So we expect this circulation to be confined within the convection zone and the most
plausible assumption is that the return flow towards the equator is located at the bottom
of the convection zone.  Choudhuri, Sch\"ussler and Dikpati [66] showed that 
a dynamo model with this type of meridional
circulation can explain the appearance of sunspots at lower latitudes with the
progress of the solar cycle.
The type of dynamo model in which the poloidal field is generated
by the Babcock--Leighton mechanism and the meridional circulation plays a crucial 
role is called the {\em flux transport dynamo model}.  It may be mentioned that the diffuse
magnetic field outside active regions migrates poleward with the solar cycle.  This is
believed to be caused by advection due to the poleward meridional circulation near
the surface [67, 68, 69, 70].  This behaviour of the poloidal field at the solar surface
automatically comes out in flux transport dynamo models. We now summarize the basic features
of this model through a cartoon.

\subsection{The whole picture}

\begin{figure}[t]
%\vspace{8cm}
\centerline{\includegraphics[width=0.7\textwidth,clip=]{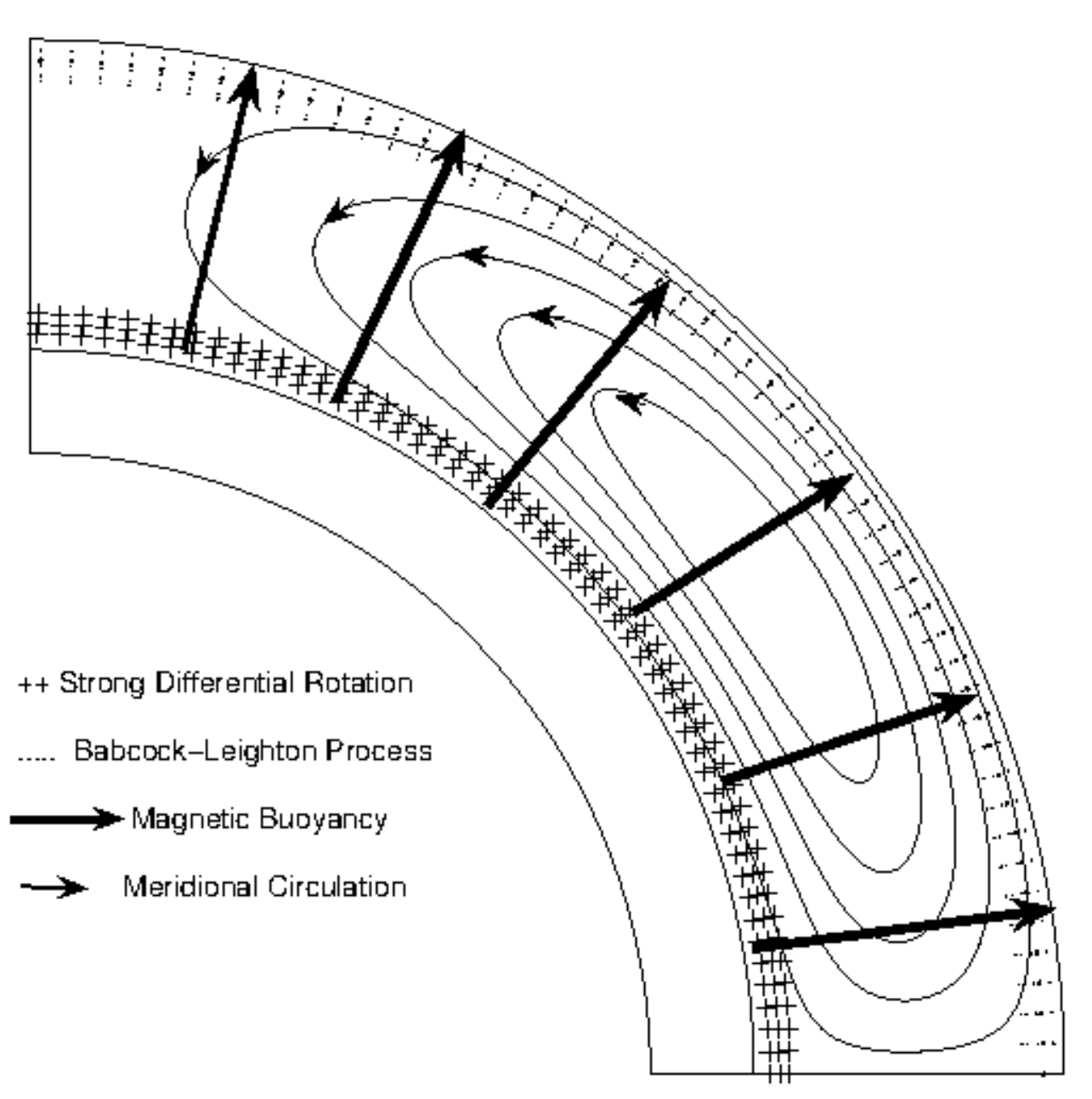}}
\caption{A schematic cartoon explaining the basic principles of the
flux transport dynamo.}
\end{figure}

Figure~12 is a cartoon encapsulating how the solar dynamo operates.
If you understand this cartoon, then you would have got the
central idea of the flux transport dynamo!
The toroidal field is produced in the tachocline by the differential
rotation stretching out the poloidal field. Then this toroidal
field rises due to magnetic buoyancy to produce bipolar sunspots
at the solar surface, where the poloidal field is generated by
the Babcock--Leighton mechanism from these bipolar sunspots.
The poloidal field so generated is carried by the meridional
circulation first to the polar region and then underneath the
surface to the tachocline to be stretched by the differential
rotation---thus completing the cycle. Since the meridional
circulation, as indicated by the streamlines sketched in Figure~12,
is equatorward at the bottom of the convection zone, the toroidal
field generated there is advected equatorward, such that sunspots
appear at increasingly lower latitudes with the progress of the
solar cycle. Although the basic idea of the flux transport dynamo
can be found in an early paper by Wang, Sheeley and Nash [71],
the first 2D models of the flux transport dynamo were constructed
in the mid-1990s by Choudhuri, Sch\"ussler and Dikpati [66] and
Durney [72].

So far we have avoided getting into equations.
For those readers who wish to see the equations, the central
equations of the flux transport dynamo theory are now shown.
In spherical coordinates, we write the
magnetic field as
$$
{\bf B} = B (r, \theta) {\bf e}_{\phi} + \nabla \times [ A
(r, \theta) {\bf e}_{\phi}],
\eqno(1)$$
where $B (r, \theta)$ is the toroidal component and $A (r, \theta)$
gives the poloidal component. We can write the velocity field
as $\vb + r \sin \theta \, \Omega (r, \theta) {\bf e}_{
\phi}$, where 
$\Omega (r, \theta)$ is the angular velocity in the interior of the
Sun and $\vb$ is the velocity of meridional circulation having components
in $r$ and $\theta$ directions.  Then the main equations telling us
how the poloidal and the toroidal fields evolve with time are
$$
\frac{\pa A}{\pa t} + \frac{1}{s}(\vf.\nabla)(s A)
= \lambda_T \left( \nabla^2 - \frac{1}{s^2} \right) A + S (r, \theta; B),
\eqno(2)$$
$$ \frac{\pa B}{\pa t} 
+ \frac{1}{r} \left[ \frac{\pa}{\pa r}
(r v_r B) + \frac{\pa}{\pa \theta}(v_{\theta} B) \right]
= \lambda_T \left( \nabla^2 - \frac{1}{s^2} \right) B 
+ s(\Bf_p.\nabla)\Omega + \frac{1}{r}\frac{d\lambda_T}{dr}
\frac{\partial}{\partial{r}}(r B), \eqno(3)$$
where $s = r \sin \theta$ and $\lambda_T$ is the turbulent
diffusivity inside the convection zone. The source term $S(r,\theta; B)$
in (2) is responsible for the generation of the poloidal field and is often
taken as $S(r, \theta; B) = \alpha B$ in many dynamo models.
We should point out that Equations (2) and (3) are
mean field equations obtained by averaging over the turbulence
in the convection zone and describe the mean behaviour of
the average magnetic field. Since Equations (2) and (3)
are coupled partial differential equations,
nothing much can be done
analytically. Our research group in IISc 
Bangalore has developed a numerical
code {\it Surya} for studying the flux transport dynamo problem
by solving these equations [73, 74].  I can
send the code {\it Surya} and a detailed guide for using it 
to anybody who sends a request to my e-mail address arnab@physics.iisc.ernet.in.

Although the flux transport dynamo has succeeded in explaining many aspects
of the observational data pertaining to solar cycles, many big uncertainties
remain. The model outlined above is of kinematic nature and the various flow
fields have to be specified in order to construct a model. While the differential
rotation has been pinned down by helioseismology [50], the nature of the meridional
circulation deep down in the convection zone remains uncertain [75, 76].
However, it is now realized that flux transport dynamos may work even with
more complicated meridional circulations than what is indicated in Figure 12 [77]. 
Several comprehensive reviews [78, 79, 80] may be recommended to readers desirous
of learning more about the current status of the flux transport dynamo
model.

\subsection{Modelling irregularities of activity cycles}

The solar cycle is only approximately periodic.  Not only the Sun, the other
stars also show irregularities in their cycle, which is evident from Figure~1.
The most notable feature of the irregularities is the grand minima, like the
Maunder minimum of the Sun during 1640--1715 when sunspots were seldom seen and several cycles
went missing.  In the time series of Ca H/K emission from many solar-like stars,
one finds evidence for grand minima.

We make a few remarks about the recent works on modelling the irregularities
of the solar cycle, since these works presumably have important implications
for stellar cycles.  Although certain aspects of the irregularities are explained
best as arising out of nonlinearities in the dynamo equations [81], the sustained
irregularities of the solar cycle are more likely caused by stochastic processes [82].
The Babcock--Leighton mechanism for generating the poloidal field depends on the tilts
of bipolar sunspot pairs.  Although the average tilt is given by Joy's law, one finds
quiet a lot of scatter around it---presumably caused by the effect of turbulence
on rising flux tubes [83]. Assuming that the randomness in the Babcock--Leighton
mechanism arising out of this scatter in tilt angles
is the main source of irregularities of the solar cycle, Choudhuri, Chatterjee
and Jiang [84] made a prediction for the strength of the present solar cycle~24 before
its onset.  This turned out to be the first successful prediction of a solar cycle
from a theoretical model, justifying the physics used in the model [85].  One important
aspect of the irregularities is the so-called Waldmeier effect: the observation that
the strengths of solar cycles are anti-correlated with their rise times. It has been possible
to explain this effect by invoking fluctuations in the meridional circulation [86].
The Ca H/K data presented by Baliunas et al. [18] show evidence for the Waldmeier
effect in several stars, indicating that these stars also must be having flux
transport dynamos inside them [86]. Choudhuri and Karak [87] developed a model of
grand minima, assuming that they are produced by combined fluctuations in the Babcock--Leighton
mechanism and the meridional circulation [88]. The theoretical efforts for modelling
irregularities have been reviewed by Choudhuri [89].

\section{The extrapolation of the flux transport dynamo to stars}

After summarizing the main features of the flux transport dynamo model, which
has been so successful in explaining different aspects of the solar cycle, we
come to the question whether this model can be extrapolated to other stars to
explain various features of their magnetic activity.  As should be clear from
the discussions of the previous section, we need to specify the differential rotation
and the meridional circulation in order to construct a model of the flux transport
dynamo.  In the case of the Sun, we have got quite a lot of information about these
large-scale flows from helioseismology. One of the main reasons behind the success
of the recent solar dynamo models is that we can use these results of helioseismology
as inputs to our solar dynamo models.  We do not have such data about the differential
rotation and the meridional circulation inside other stars---except some data about 
differential rotation at the surface for a few stars [35]. So the first big hurdle
for constructing models of stellar dynamos is that we have to figure out the
nature of these large-scale flows inside different stars from purely theoretical
considerations.

\subsection{Large-scale flows inside stars}

In a pioneering study in 1963, Kippenhahn [90] showed that an anisotropic viscosity
can give rise to large-scale flows inside a rotating star.  If the radial viscosity
is larger, then that causes slower rotation at the equatorial region.  In order to
have faster rotation near the equator, which is the case for the Sun, we need to
have larger horizontal viscosity.  Within the Sun's convection zone, viscosity is
mainly provided by turbulence, compared to which the molecular viscosity is
negligible. The turbulent viscosity within the convection zone is certainly
expected to be anisotropic due to two factors: (i) gravity makes the radial
direction special; and (ii) rotation makes the polar direction special. To get a
physical picture of how the large-scale flows are induced by turbulent transport
mechanisms inside the convection zone of a rotating stars, see the very clear reviews by
Kitchatinov [91, 92] on this complex subject.

In order to compute large-scale flows inside stars, we can follow one of the
two possible approaches.  The first approach is to do a direct numerical simulation
of convection in a rotating star from first principles.  Such simulations have
shown the occurrence of differential rotation and meridional circulation.
The second approach is to first calculate the various components of turbulent viscosity
from a mixing length theory of convective turbulence and then to use these in a
mean field model of stellar hydrodynamics to compute the large-scale flows inside 
the stars.  This second approach was pioneered by Kitchatinov and R\"udiger [93].

Once we have the large-scale flows inside a star, we can substitute these in the
equations of the flux transport dynamo and obtain a model of the stellar dynamo.
Jouve, Brown and Brun [94] constructed stellar dynamo models by following the first
approach of computing the large-scale flows through direct numerical simulations.
Constructing such models is computationally demanding and it is difficult to explore
the parameter space extensively by following this approach.  On the other hand,
Karak, Kitchatinov and Choudhuri ([95], hereafter KKC) followed the second approach of computing
the large-scale flows from the mean field model of Kitchatinov and Olemskoy [96]
and then constructing stellar dynamo models.  In this approach, it is possible
to explore the parameter space more extensively.

\subsection{Comparing stellar dynamo models with observations}

Figure~13 shows the differential rotations computed by KKC [95] for stars
of mass $1 M_{\odot}$ rotating with different rotation periods. For the slowly rotating case
with a period of 30 days (close to the solar rotation period of 27 days), the angular
velocity is constant over cones in the convection zone, similar to what is found for
the Sun from helioseismology.  On the other hand, for the rapidly rotating case with
a period of 1 day, the angular velocity tends to be constant over cylinders. The meridional
circulation is also computed and is found to be weaker for faster rotators. It may
be noted that, when the rotation is very slow and its effect on convection negligible,
the radial part of viscosity becomes dominant due to the primarily up-down nature
of the convective motions and the differential rotation changes
over to an anti-solar pattern with slower rotation near the equator, as confirmed
by recent simulations [97]. This case is not covered in Figure~13.

\begin{figure}[t]
%\vspace{8cm}
\centerline{\includegraphics[width=0.9\textwidth,clip=]{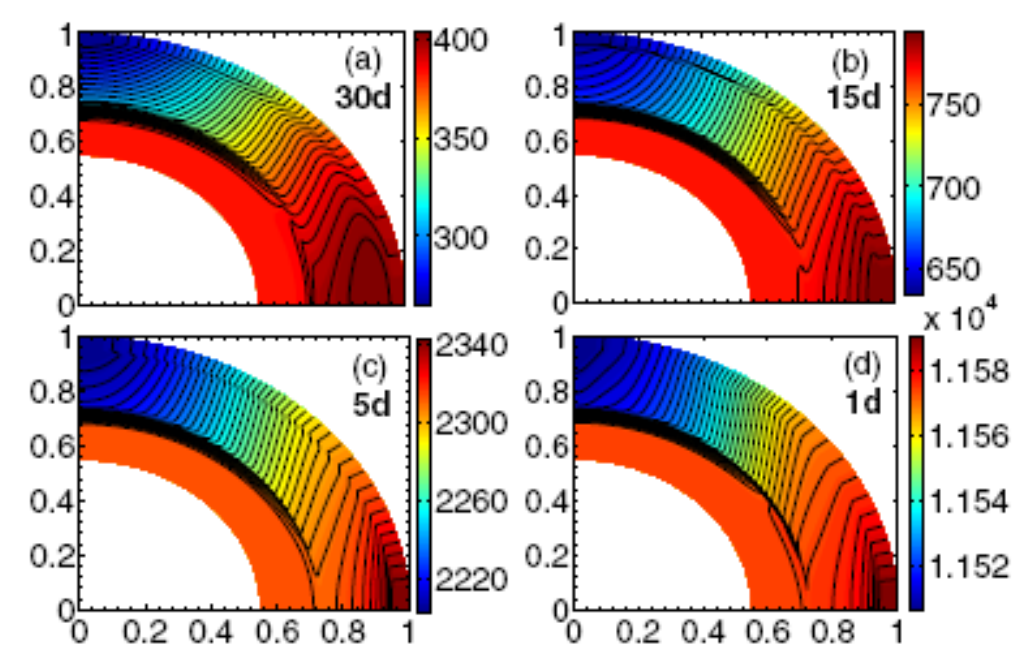}}
\caption{The angular velocity profiles of $1 M_{\odot}$ stars rotating with rotation
periods of (a) 30 days, (b) 15 days, (c) 5 days, and (d) 1 day, as computed
from a mean field model.  From Karak, Kitchatinov and Choudhuri (KKC) [95].}
\end{figure}

KKC [95] constructed dynamo models of $1 M_{\odot}$ mass stars rotating with
different angular speeds by inserting the differential rotation and the meridional
circulation computed from the mean field model into the dynamo equations (2) and (3). In order to
make sure that the dynamo solutions do not grow indefinitely, it is necessary to
include a quenching.  They took the source function appearing in Equation (2) to have the
form
$$S (r,\theta; B) = \frac{\alpha(r, \theta)}{1 + (B(r_t, \theta)/B_0)^2} B(r_t. \theta),
\eqno(4)$$
where $B(r_t, \theta)$ is the toroidal field at the bottom of the convection zone ($r = r_t$)
and the coefficient $\alpha( r, \theta)$ is assumed to be concentrated near the
surface, to account for the Babcock--Leighton mechanism in which the toroidal field
from the bottom of the convection zone rises to the surface to produce a poloidal
field at the surface.  This coefficient, although denoted by the symbol $\alpha$,
has a physical origin completely different from the traditional $\alpha$-effect. 
The quenching factor appearing in the denominator ensures that
the source term becomes very small when the toroidal field $B(r_t,\theta)$ is much
larger than $B_0$.  Hence the dynamo is found to saturate with $B(r_t,\theta)$ hovering 
around a value not much larger than $B_0$. The total toroidal flux through the convection
zone can be written as $f B_0 R_{\odot}^2$. KKC calculated $f$ from their dynamo model
and the mean $f_m$ of its unsigned value averaged over the cycle was
taken as a measure of the toroidal flux generated in a particular
situation.  Since Ca H/K or X-ray emission presumably arises from energy generated due
to magnetic reconnection between two flux systems, we may naively expect these emissions
to be proportional to $f_m^2$. Figure~14 taken from KKC shows how $f_m^2$ varies with the
Rossby number in the theoretical model. If we assume that there is some mechanism
which saturates the Babcock--Leighton mechanism for fast rotations, then we get a
theoretical curve which agrees with Figure~3 (Ca H/K emission plot) or Figure~6 (X-ray
emission plot) remarkably well.

\begin{figure}[t]
%\vspace{8cm}
\centerline{\includegraphics[width=0.7\textwidth,clip=]{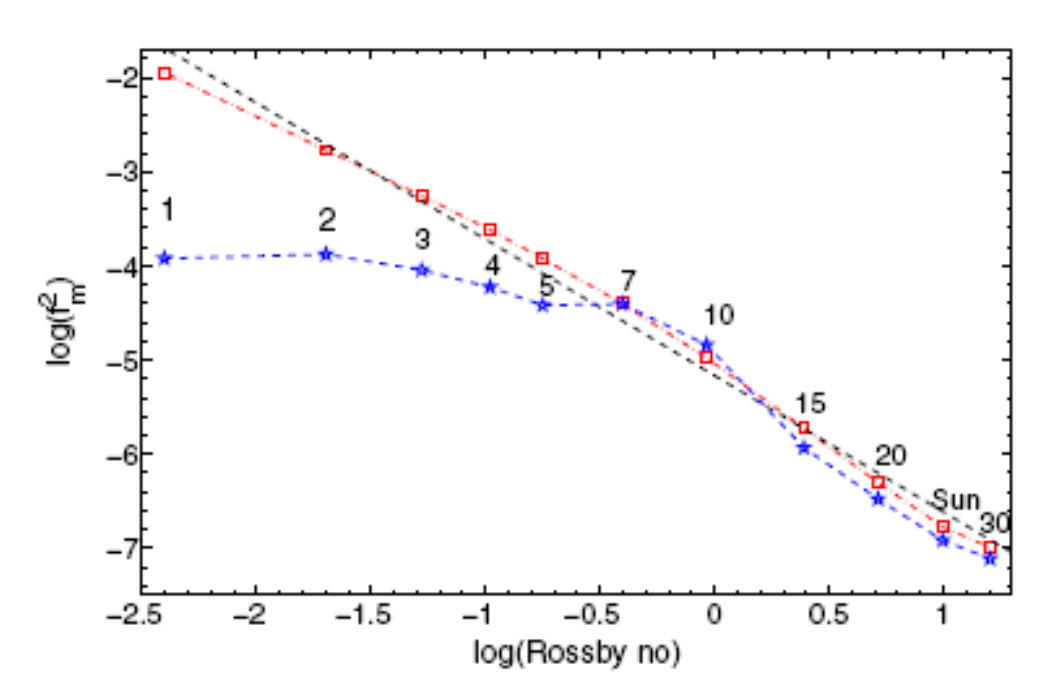}}
\caption{The theoretically computed $f_m^2$ as a function of the Rossby number. The blue curve
corresponds to the case in which we assume a saturation of the Babcock--Leighton mechanism
for fast rotators, whereas the red curve corresponds to the case without such a saturation.
From KKC [95].}
\end{figure}

While we were happy that we (KKC) could model the increase of emission with lower Rossby
number, we failed to explain the observed increase of cycle period with rotation period,
as indicated in Figure~4.  We reproduce the theoretical plot from KKC in Figure~15, showing
the increase of the cycle period with decreasing rotation period, contradicting the observational
data.  This results from the fact that the meridional circulation becomes weaker in faster
rotating stars according to the mean field hydrodynamic model which we had used in order to calculate
the meridional circulation.  Since the period of the flux transport dynamo depends on
the time scale of meridional circulation, a faster rotating star (with shorter rotation
period) gives rise to weaker meridional circulation and therefore longer cycle period.
Jouve, Brown and Brun [94], who computed the meridional circulation from direct numerical
simulations, also found the same difficulty.  An intriguing question is whether there is
a flaw in our understanding of the meridional circulation and whether it could be stronger
for more highly rotating stars, which would solve this problem.  Interestingly, in the
traditional $\alpha \Omega$ dynamo model, the cycle frequency goes as the square root
of ($\alpha \times$ gradient of angular velocity) (see [49], p.\ 360). 
While the flux transport dynamo model fails
to explain the observed relation between the rotation period and the cycle period, this
relation can be explained easily in the traditional $\alpha \Omega$ 
dynamo model on assuming that $\alpha$ and/or the gradient of angular velocity 
increase with increasing rotation frequency, as pointed out by Noyes, Weiss and Vaughan [22]. This
raises the question whether the nature of the stellar dynamo changes in stars rotating very
fast for which the meridional circulation will be weak and we may have an $\alpha \Omega$ dynamo
instead of a flux transport dynamo.  These questions remain to be addressed by future research.

It should be clear from our brief discussion of the physics of the flux transport dynamo
in \S~3 that the bottom of the convection zone plays an important role in the dynamo process.
It is there that the toroidal field is produced by the strong differential rotation and then
a part of it remains stored below the bottom where the magnetic buoyancy is suppressed.
Now, stars having mass less than about $0.4 M_{\odot}$ (of spectral type later than
M3--3.5) are supposed to be fully convective, without a bottom below which the toroidal
field can be stored.  Whether the usual flux transport dynamo can operate in such a star
is an important question.  Recently, Wright and Drake [98] pointed out that X-ray emission from
some fully convective stars satisfy the relation between X-ray luminosity and Rossby number
that we see in Figure~6, suggesting that these stars also have dynamos similar to other late-type
stars.  Although there have been some works on dynamo action in fully convective stars [99, 100, 101],
our understanding of this subject is still very incomplete.

\begin{figure}[t]
%\vspace{8cm}
\centerline{\includegraphics[width=0.6\textwidth,clip=]{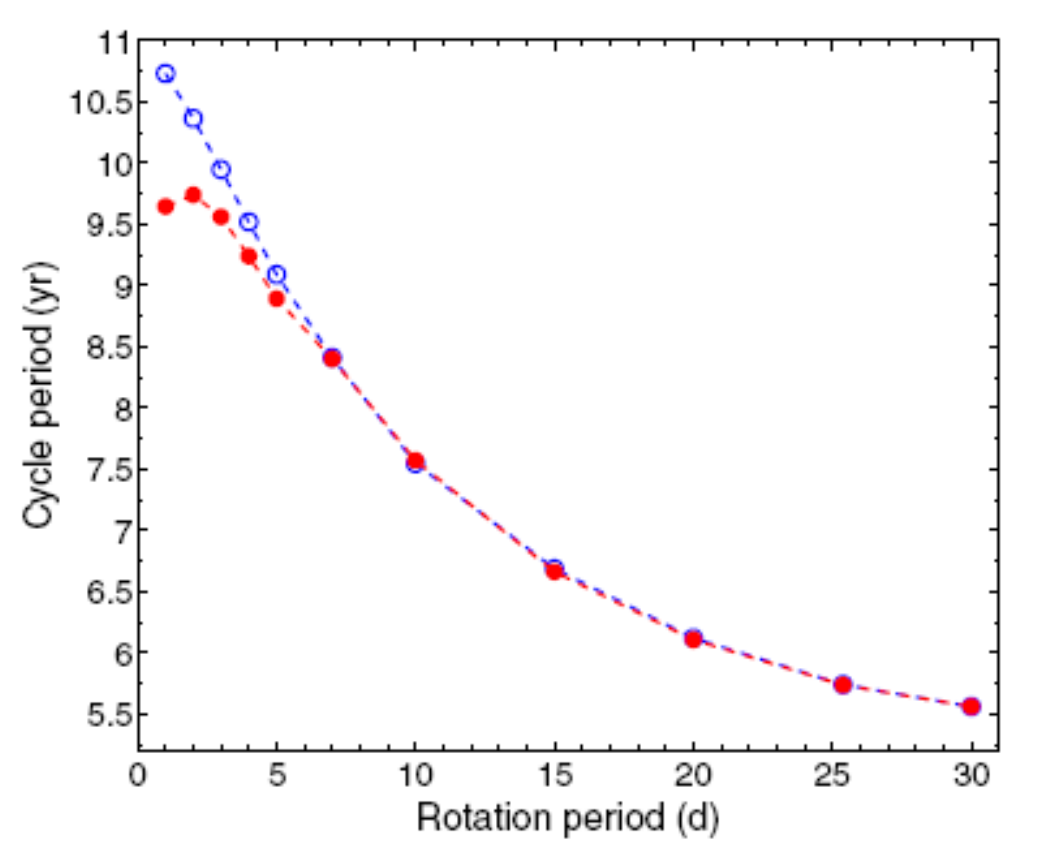}}
\caption{The variation of the cycle period with the rotation period according to the
theoretical model of KKC [95].}
\end{figure}

\subsection{Large sunspots and strong flares}

As mentioned in \S~2.3 and 2.4, we have evidence for starspots much larger than the
largest sunspots and stellar superflares much stronger than the strongest solar flares.
One important question is whether such large starspots and strong stellar flares require
physical mechanisms different from what are operative in the Sun, indicating that
the nature of the dynamo also may be somewhat different.  It is difficult to answer
this question at the present time because we have very little understanding of what
determines the sizes of sunspots or fluxes in active regions.  Since we believe that
some amount of magnetic flux broken from the toroidal flux system stored at the bottom of 
the solar convection zone rises to the surface to produce sunspots and active regions,
presumably their sizes depend on the nature of the instabilities that break up the toroidal
flux at the bottom of the convection zone (see the review by Fan [102]).  
Our understanding of the storage and breakup process
of the toroidal flux is very poor at present.  The best we can do is to try to estimate
maximum possible sizes of sunspots based on some `reasonable' assumptions.

As we pointed out in \S~2.3, large starspots in rapidly rotating stars are often
found near the polar region.  Sch\"ussler and Solanki [103] provided an explanation
for this by extrapolating the results of Choudhuri and Gilman [56], who studied the effect
of the Coriolis force on magnetic buoyancy (see also[104]).  As discussed in \S~3.2, Choudhuri and
Gilman [56] found that, when the Coriolis force wins over magnetic buoyancy, the magnetic flux
rising due to magnetic buoyancy is diverted by the Coriolis force to rise parallel to
the rotation axis and emerge at high latitudes.  Presumably this is what happens in
rapidly rotating stars, causing the rising flux to emerge at polar latitudes to create
polar starspots. Isik, Schmitt and Sch\"ussler [105] combined a dynamo model with magnetic
buoyancy calculations to study the distribution of starspots over the stellar surface.

\begin{figure}[t]
%\vspace{8cm}
\centerline{\includegraphics[width=0.65\textwidth,clip=]{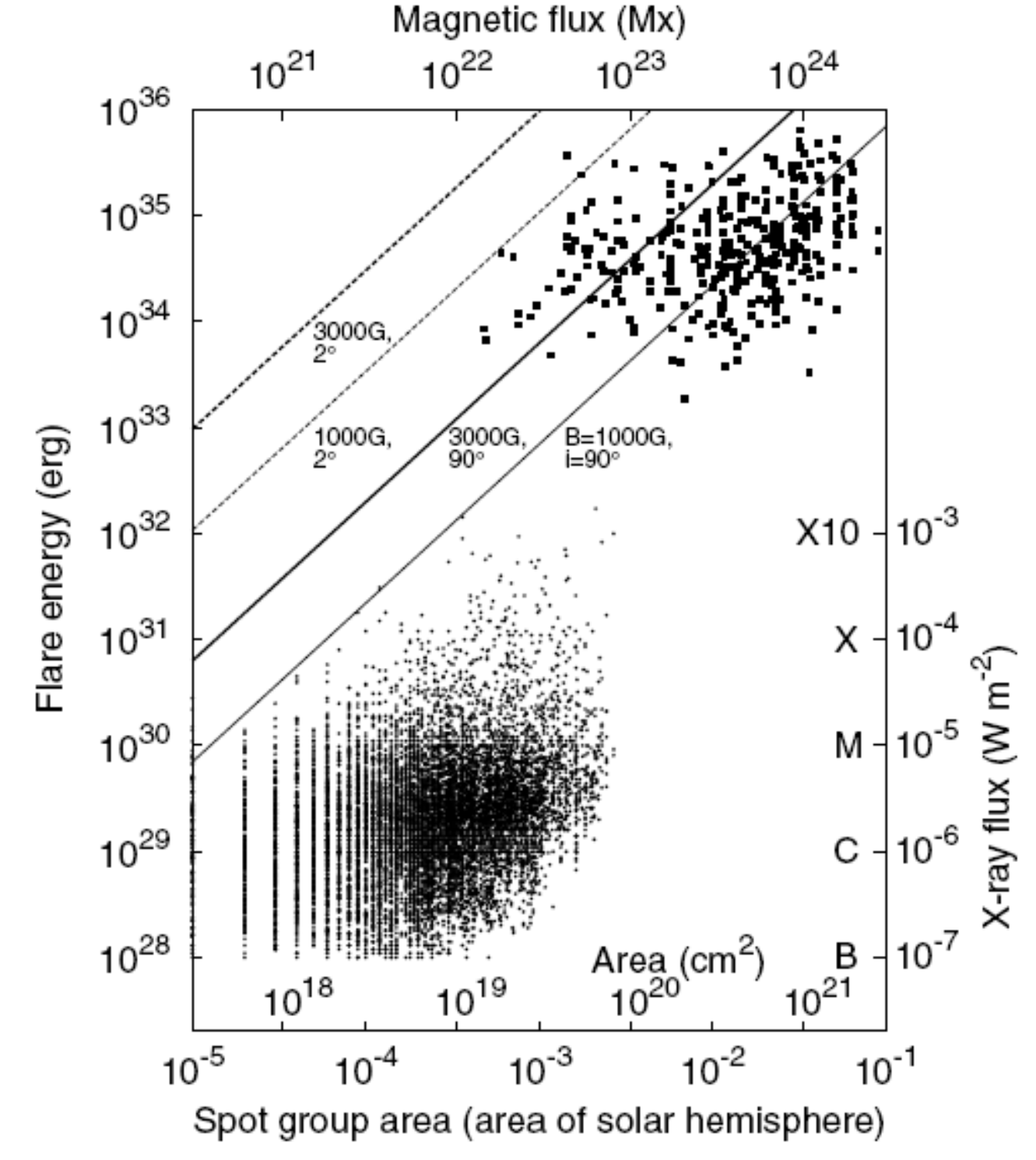}}
\caption{Flare energies plotted against areas (or equivalently magnetic fluxes) of the active regions
with which the flares were associated.  Both solar flares (lower left) and stellar superflares
(upper right) are shown.  From Shibata et al.\ [40].}
\end{figure}

Since we know of stellar flares much stronger than the strongest solar flares recorded
so far, we now come to the question whether these require different physics or whether it
is possible for such strong flares to occur on the Sun also.  Since the largest known solar
flares (of energy of order $10^{32}$ erg) have been known to cause serious disruptions in
human activities (especially on electrical and electronic appliances), significantly
stronger solar flares will certainly cause havoc and disrupt our lives in major ways.
Figure~16 taken from Shibata et al.\ [40] plots the energy of flares against the spot
group area (or, equivalently, magnetic flux) with which these flares have been
associated.  Both solar flares (lower left region of the figure) and stellar flares
(upper right region) are shown.  Note that only stellar flares much more energetic
than solar flares can be observed by us.  However, it appears from the figure that there
might be a continuity between the solar flares and stellar superflares.  Figure~16 seems to
suggest that superflares of energy $10^{35}$ erg would be associated with 
spot groups having magnetic flux of about
$10^{24}$ Mx, whereas the largest sunspots carry flux not more than $10^{23}$ Mx.
Whether such superflares can occur on the Sun then basically hinges on the question
whether we can have active regions with such flux.  Pushing the various parameters 
connected with the dynamo generation of the toroidal field to
their extreme values, Shibata et al.\ [40] concluded that this is not entirely impossible,
although we are not sure whether the extreme values assumed in the estimate
are completely justified.  If these are justified, then we have to conclude that the occurrence
of such superflares may not require dynamo action qualitatively different from the solar
dynamo and, at the same time, we cannot rule out such superflares on our Sun occurring
with extremely low frequency---such as a superflare of energy $10^{35}$ erg in 5000
years as estimated by Maehara et al.\ [39]. It is also possible that large fluctuations
in dynamo parameters may push a star to an extreme temporary phase when superflares
are more likely [106]. 

\section{Conclusion}

In the last few years a huge amount of data about magnetic activity of solar-like stars 
have come.  We have also witnessed remarkable developments in solar dynamo modelling.
To the best of our knowledge, this is the first attempt to review the question whether the current 
solar dynamo models can be extrapolated to model magnetic activities of other solar-like
stars. 

Stellar magnetism is a vast field---magnetic fields playing important roles in the star
formation process as well as in the final phase of stellar collapse (see [107, 108]).  It may be
mentioned that a modified version of our solar dynamo code has been used to study the role of
magnetic fields in accreting neutron stars [109, 110]. In this review, however, we restrict ourselves
only to late-type solar-like stars in their main sequence, having convection zones just below
their surfaces. Such stars are found to have activity cycles, coronae, spots and flares just
like the Sun.  Very intriguingly, some of them have spots much larger than sunspots and
flares much stronger than solar flares.

A particularly successful model of the solar cycle is the flux transport dynamo model, in
which the poloidal field is generated by the Babcock--Leighton mechanism and the
meridional circulation plays an important role.  A very pertinent question is whether
other solar-like stars also have such flux transport dynamos operating inside them. We
have seen that such dynamo models can explain the relation between the activity-related
emission (in Ca H/K or in X-ray) and the Rossby number, giving us confidence that we
are probably on the right track. We have pointed out that Ca H/K emission of several
stars indicate the Waldmeier effect, which has been explained as arising out of the
fluctuations of the meridional circulation [86]. This presumably indicates that these stars
also have meridional circulation with fluctuations.  However, we should keep in mind
that the flux transport dynamo model failed to explain the observed relation between 
the rotation period and the cycle period, which could be explained easily by the
earlier $\alpha \Omega$ dynamo model [22].  This raises the question whether the meridional
circulation becomes very weak in rapidly rotating stars and the dynamo changes over
to an $\alpha \Omega$ dynamo from a flux transport dynamo. In this connection, note
also that the cycle of chromospheric Ca H/K emission may not always indicate the
magnetic cycle [111]. We know that for rapidly
rotating stars polar regions will dominate the activity, as the Coriolis force will deflect
the magnetic flux rising due to magnetic buoyancy to polar regions [56, 103]. However, we are
not sure whether very large starspots and very strong stellar superflares can be explained
by extrapolating solar models or some different kinds of dynamo models are needed. The
accompanying question is whether we should expect to see much larger sunspots or
much stronger solar flares than the ones we have so far seen.  There is no doubt that many
solar-like stars have flux transport dynamos like the Sun.  But is this some kind of
universal dynamo model that can account for magnetic activity of all stars---including
those which have very large spots and very strong flares? Further research is definitely
needed to answer this question.

While summarizing the flux transport dynamo model in \S~3, we have restricted
ourselves to 2D kinematic models.  While these models have been very successful in
explaining different aspects of solar activity, these models have obvious limitations
and there are efforts under way to go beyond these simple models. For example,
magnetic buoyancy leading to the Babcock--Leighton mechanism is an inherently
3D process and cannot be captured realistically in 2D models [112]. There are now
attempts of constructing 3D kinematic models [113, 114, 115].   However, we ultimately need
to go beyond kinematic models and develop fully dynamical 3D models.  Some
exploratory studies have produced striking initial results [116, 117].  There is no doubt that
such developments will have a big impact on stellar dynamo research in future
and this will remain a very active research field for years to come.

\bigskip

{\em Acknowledgements.}  I thank Peng-Fei Chen for inviting me to write this review.
Suggestions from Leonid Kitchatinov, Karel Schrijver, Klaus Strassmeier and
Sharanya Sur helped in improving an earlier version of the review.
I am grateful to Gopal Hazra for discussions and help in preparing the
manuscript. Partial support for this work was provided by the J C Bose Fellowship
awarded by the Department of Science and Technology, Government of India.

\def\apj{{Astrophys.\ J.}}
\def\mnras{{Mon.\ Notic.\ Roy.\ Astron.\ Soc.}}
\def\sol{{Solar Phys.}}
\def\aa{{Astron.\ Astrophys.}}
\def\gafd{{\it Geophys.\ Astrophys.\ Fluid Dyn.}}

%\end{document}


\begin{thebibliography}{}
 \label{reflist}

\bibitem[1]{maoz}
  D. Maoz, {\em Astrophysics in a Nutshell}, Princeton University Press (2007).
  
\bibitem[2]{cho10a}
  A. R. Choudhuri, {\em Astrophysics for Physicists}, Cambridge University Press (2010).
  
\bibitem[3]{gal}
  G. Galileo, {\em History and Demonstration Concerning Sunspots} (1613).
  
\bibitem[4]{hale08}
  G. E. Hale,   \apj\ {\bf 28}, 315 (1908).

\bibitem[5]{sch44}
  S. H. Schwabe,   {Astron.\ Nachr.} {\bf 21}, 233 (1844).

\bibitem[6]{edl}
  B. Edl\'en, Zeits. f. Astroph. {\bf 22}, 30 (1943).
  
\bibitem[7]{low}
  B. C. Low, { Sciene China Physics, Mechanics \& Astronomy} {\bf 58}, 5626 (2015).
  
\bibitem[8]{car59}
  R. C. Carrington,  \mnras\ {\bf 20}, 13 (1859).
  
\bibitem[9]{sch00}
  C. J. Schrijver, and C. Zwaan, {\em Solar and Stellar Magnetic Activity}, Cambridge University Press (2000).
  
\bibitem[10]{thomas08}
  J. H. Thomas, and N. O. Weiss, {\em Sunspots and Starspots}, Cambridge University Press (2008).
 
\bibitem[11]{brun15}
  A. S. Brun, R. A. García, G. Houdek, D. Nandy, and M. Pinsonneault, {Space Sci. Rev.} {\bf 196}, 303 (2015).
  
\bibitem[12]{iau273}
  D. P. Choudhary, and K. G. Strassmeier (editors), {\em Physics of Sun and Star Spots --- IAU Symposium 273}, Cambridge University Press (2011).
  
\bibitem[13]{iau286}
  C. H. Mandrini, and D. F. Webb (editors), {\em Comparative Magnetc Minima: Characterizing Quiet Times in the Sun and 
  Stars --- IAU Symposium 286}, Cambridge University Press (2012).

\bibitem[14]{eber}
  G. Eberhard, and K. Schwarzschild, \apj\ {\bf 38}, 292 (1913).

\bibitem[15]{wil57}
  O. C. Wilson, and M. K. V. Bappu, \apj\ {\bf 125}, 661 (1957).
  
\bibitem[16]{hall}
  J. C. Hall, Living Rev. Solar Phys. {\bf 5}, 2 (2008).
  
\bibitem[17]{wil76}
  O. C. Wilson, \apj\ {\bf 226}, 379 (1976).  
  
\bibitem[18]{bal95}
  S. L. Baliunas, et al., \apj\ {\bf  438}, 269 (1995).
  
\bibitem[19]{vau}  
  A. H. Vaughan, and G. W. Preston, Publ. Astron. Soc. Pacific {\bf 92}, 385 (1980).
  
\bibitem[20]{noyes84a}
  R. W. Noyes, L. W. Hartmann, S. L. Baliunas, D. K. Duncan, and A. H. Vaughan, \apj\ {\bf 279}, 763 (1984).

\bibitem[21]{skew}
  A. Skumanich, \apj\ {\bf 171}, 565 (1972).

\bibitem[22]{noyes84b}
  R. W. Noyes, N. O. Weiss, and A. H. Vaughan, \apj\ {\bf 287}, 769 (1984). 
  
\bibitem[23]{saar99}
  S. H. Saar, and A. Brandenburg, \apj\ {\bf 524}, 295 (1999).
  
\bibitem[24]{palla}  
  R. Pallavicini, L. Golub, R. Rosner, G. S. Vaiana, T. Ayres, and J. L. Linsky, \apj\ {\bf 248}, 279 (1981).

\bibitem[25]{dob}
  C. J. Schrijver, A. K. Dobson, and R. R. Radick, \apj\ {\bf 258}, 432 (1992).
  
\bibitem[26]{pizzo}  
  N. Pizzolato, A. Maggio, G. Micela, S. Sciortino, and P. Ventura, \aa\ {\bf 397}, 147 (2003).
  
\bibitem[27]{wright}  
  N. J. Wright, J. J. Drake, E. E. Mamajek, and G. W. Henry, \apj\ {\bf 743}, 48 (2011).
  
\bibitem[28]{vogt}
  S. S. Vogt, and G. D. Penrod, Publ. Astron. Soc. Pacific {\bf 95}, 565 (1983).
  
\bibitem[29]{strass99}
  K. G. Strassmeier, \aa\ {\bf 347}, 225 (1999).
  
\bibitem[30]{ku15}  
  A. K\"unstler, T. A. Carroll, and K. G. Strassmeier, \aa\ {bf 578}, 101 (2015).
  
\bibitem[31]{donati}
  J.-F. Donati, M. Semel, B. D. Carter, D. E. Rees, and A. Collier Cameron, \mnras\ {\bf 291}, 658 (1997).
  
\bibitem[32]{bedyu05}
  S. V. Berdyugina, Living Rev. Solar Phys. {\bf 2}, 8 (2005).
  
\bibitem[33]{strass05}
  K. G. Strassmeier, Ann. Rev. Astron. Astrophys. {\bf 17}, 251 (2009). 
  
\bibitem[34]{Rei}
  A. Reiners, Living Rev. Solar Phys. {\bf 8}, 1 (2012).

\bibitem[35]{barnes}  
  J. R. Barnes, A. Collier Cameron, J.-F. Donati, D. J. James, S. C. Marsden, and P. Petit, \mnras\ {\bf 357}, L1 (2005).

\bibitem[36]{berdy98}
  S. V. Berdyugina, and I. Tuominen, \aa\ {\bf 336}, L25 (1998). 

\bibitem[37]{els}  
  D. Elstner, and H. Korhonen, Astron. Nachr. {\bf 326}, 278 (2005). 
  
\bibitem[38]{sche00}  
  B. E. Schaefer, J. R. King, and C. P. Deliyannis, \apj\ {\bf 529}, 1026 (2000).
  
\bibitem[39]{mae}  
  H. Maehara, et al., Nature {\bf 485}, 478 (2012).
  
\bibitem[40]{shib}  
  K. Shibata, et al., Publ. Astron. Soc. Japan {\bf 65}, 49 (2013).
  
\bibitem[41]{tsu}  
  B. T. Tsurutani, W. D. Gonzalez, G. S. Lakhina, and S. Alex, J. Geophys. Res. A {\bf 108}, 1268 (2003).
 
\bibitem[42]{hale19}
  G. E. Hale, F. Ellerman, S. B. Nicholson and A. H. Joy,
  \apj\ {\bf 49}, 153 (1919).

\bibitem[43]{bab55}
  H. W. Babcock and H. D. Babcock,  \apj\ {\bf 121}, 349 (1955).
  
\bibitem[44]{tsu08}
  S. Tsuneta, et al, \apj\ {\bf 688}, 1374 (2008).
  
\bibitem[45]{cha52}
  S. Chandrasekhar,  {Phil.\ Mag.\ (7)} {\bf 43}, 501 (1952).

\bibitem[46]{wei81}
  N. O. Weiss,  {J.\ Fluid Mech.} {\bf 108}, 247 (1981).
  
\bibitem[47]{par55b}
  E. N. Parker,  \apj\ {\bf 122}, 293 (1955).  
  
\bibitem[48]{hath}
  D. H. Hathaway, Living Rev. Solar Phys. {\bf 7}, 1 (2010).

\bibitem[49]{cho98}
  A. R. Choudhuri,  {\it The Physics of Fluids and Plasmas:
  An Introduction for Astrophysicists}, Cambridge University Press (1998).  
  
\bibitem[50]{sch98}
  J. Schou, et al., \apj\ {\bf 505}, 390 (1998).
  
\bibitem[51]{par55a}
  E. N. Parker,  \apj\ {\bf 121}, 491 (1955).
  
\bibitem[52]{par75}
  E. N. Parker,  \apj\ {\bf 198}, 205 (1975). 
  
\bibitem[53]{moreno}
  F. Moreno-Insertis, \aa\ {\bf 122}, 241 (1983).
  
\bibitem[54]{spr81}
  H. C. Spruit,  \aa\ {\bf 98}, 155 (1981). 

\bibitem[55]{chou90}
  A. R. Choudhuri,  \aa\ {\bf 239}, 335 (1990).

\bibitem[56]{cho87}
  A. R. Choudhuri, and P. A. Gilman,  \apj\ {\bf 316}, 788 (1987).

\bibitem[57]{cho89}
  A. R. Choudhuri,  \sol\ {\bf 123}, 217 (1989).

\bibitem[58]{sil93}
  S. D'Silva, and A. R. Choudhuri,  \aa\ {\bf 272}, 621 (1993).

\bibitem[59]{fan93}
  Y. Fan, G. H. Fisher, and E. E. DeLuca,  \apj\ {\bf 405}, 390 (1993).

\bibitem[60]{cal95}
  P. Caligari, F. Moreno-Insertis,  and M. Sch\"ussler,  \apj\
  {\bf 441}, 886 (1995).

\bibitem[61]{cho90}
  A. R. Choudhuri, and S. D'Silva, \aa\ {\bf 239}, 326 (1990).

\bibitem[62]{sil91}
  S. Z. D'Silva and A. R. Choudhuri, \sol\ {\bf 136}, 201 (1991).
  
\bibitem[63]{ste66}
  M. Steenbeck, F. Krause, and K.-H. R\"adler,  {Z.\ Naturforsch.} {\bf 21a}, 1285 (1966).
  
\bibitem[64]{bab61}
  H. W. Babcock,  \apj\ {\bf 133}, 572 (1961).

\bibitem[65]{lei69}
  R. B. Leighton,  \apj\ {\bf 156}, 1 (1969).
  
\bibitem[66]{cho95}
  A. R. Choudhuri, M Sch\"ussler, and M Dikpati,  \aa\
  {\bf 303}, L29 (1995).
  
\bibitem[67]{wan89}
  Y.-M. Wang, A. G. Nash and N. R. Sheeley,  \apj\
  {\bf 347}, 529 (1989).
  
\bibitem[68]{dik94}
  M. Dikpati, and A. R. Choudhuri, \aa\ {\bf 291}, 975 (1994).
  
\bibitem[69]{dik95}
  M. Dikpati, and A. R. Choudhuri, \sol\ {\bf 161}, 9 (1995).
    
\bibitem[70]{dik99}  
  A. R. Choudhuri, and M. Dikpati, \sol\ {\bf 184}, 61 (1999).
  
\bibitem[71]{wsn89}
  Y.-M. Wang, N. R. Sheeley, and A. G. Nash, \apj\ {\bf 383}, 431 (1991).
  
\bibitem[72]{dur95}
  B. R. Durney,  \sol\ {\bf 160}, 213 (1995).  
  
\bibitem[73]{nan02}
  D. Nandy, and A. R. Choudhuri,  {Science} {\bf 296}, 1671 (2002).

\bibitem[74]{cha04}
  P. Chatterjee, D. Nandy, and A. R. Choudhuri,  \aa\ {\bf 427}, 1019 (2004).

\bibitem[75]{zhao}
  J. Zhao, R. S. Bogart, A. G. Kosovichev, T. L. Duvall, and T. Hartlep, \apj\ {\bf 774}, L29 (2013).
  
\bibitem[76]{raja}  
  S. P. Rajaguru, and H. M. Antia, \apj\ {\bf 813}, 114 (2015).
  
\bibitem[77]{haz14}
  G. Hazra, B. B. Karak, and A. R. Choudhuri, \apj\ {\bf 782}, 93 (2014).
  
\bibitem[78]{cho11}
  A. R. Choudhuri, Pramana {\bf 77}, 77 (2011).
  
\bibitem[79]{char14}  
  P. Charbonneau, Ann. Rev. Astron. Astrophys. {\bf 52}, 251 (2014).
  
\bibitem[80]{kar14}  
  B. B. Karak, J. Jiang, M. S. Miesch, P. Charbonneau, and A. R. Choudhuri, Space Sci. Rev. {\bf 186}, 561 (2014).
  
\bibitem[81]{cha06}
  P. Charbonneau, G. Beaubien, and C. St-Jean, \apj\ {\bf 658}, 657 (2007).
    
\bibitem[82]{cho92}
  A. R. Choudhuri, \aa\ {\bf 253}, 277 (1992).  
  
\bibitem[83]{lon02}
  D. Longcope, and A. R. Choudhuri,  \sol\ {\bf 205}, 63 (2002).  
  
\bibitem[84]{cho07}
  A. R. Choudhuri, P. Chatterjee, and J. Jiang, {Phys.\ Rev.\ Lett.} 
  {\bf 98}, 131103 (2007).
  
\bibitem[85]{jia07}
  J. Jiang, P. Chatterjee, and A. R. Choudhuri, \mnras\ {\bf 381}, 1527 (2007).
  
\bibitem[86]{kar11}
  B. B. Karak, and A. R. Choudhuri, \mnras\ {\bf 410}, 1503 (2011). 
  
\bibitem[87]{cho12}
  A. R. Choudhuri, and B. B. Karak, Phys. Rev. Lett. {\bf 109}, 171103 (2012).
  
\bibitem[88]{kar13}
  B. B. Karak, and A. R. Choudhuri, Res. Astron. Astrophys. {\bf 13}, 1339 (2013).  
  
\bibitem[89]{cho14}
  A. R. Choudhuri, Indian J. Phys. {\bf 88}, 877 (2014).
  
\bibitem[90]{kip}
  R. Kippenhahn, \apj\ {\bf 137}, 664 (1963).
    
\bibitem[91]{kit11}
  L. L.Kitchatinov, in {\em Proceedings of the First Asia-Pacific Solar Physics Meeting} (eds.: A. R. Choudhuri and
  D. Banerjee), p.\ 71 (2011).
  
\bibitem[92]{kit11}
  L. L.Kitchatinov, in {\em Solar and Astrophysical Dynamos and Magnetic Activity --- IAU Symposium 294} (eds.: A. G. Kosovichev,
  E. de Gouveia Dal Pino and Y. Yan), p.\ 399 (2013).
  
\bibitem[93]{kit95}
  L. L. Kitchatinov, and G. R\"udiger, \aa\ {\bf 299}, 446 (1995).
    
\bibitem[94]{jou}  
  L. Jouve, B. P. Brown, and A. S. Brun, \aa\ {\bf 509}, 32 (2010).
  
\bibitem[95]{kar14}
  B. B. Karak, L. L. Kitchatinov, and A. R. Choudhuri, \apj\ {\bf 791}, 59 (2014). 
  
\bibitem[96]{kit11a}
  L. L. Kitchatinov, and S. V. Olemskoy, \mnras\ {\bf 411}, 1059 (2011).  
  
\bibitem[97]{kar15}  
  B. B. Karak, P. J. K\"apyl\"a, M. J. K\"apyl\"a, A. Brandenburg, N. Olspert, and J. Pelt, \aa\ {\bf 576}, 26 (2015).
  
\bibitem[98]{wri}
  N. J. Wright, and J. J. Drake, Nature {\bf 535}, 526 (2016).
  
\bibitem[99]{dur93}
  B. R. Durney, D. S. De Young, and I. W. Roxburgh, \sol\ {\bf 145}, 207 (1993).
  
\bibitem[100]{dob}
  W. Dobler, M. Stix, and A. Brandenburg, \apj\ {\bf 638}, 336 (2006).
  
\bibitem[101]{brown}
  M.K. Browning, \apj\ {\bf 676}, 1262 (2008).
  
\bibitem[102]{fan04}
  Y. Fan, Living Rev. Solar Phys. {\bf 1}, 1 (2004).
  
\bibitem[103]{sch92}
  M. Sch\"ussler, and S. K. Solanki, \aa\ {\bf 264}, L13 (1992).
  
\bibitem[104]{sch96}
  M. Sch\"ussler, P. Caligari, A. Ferriz-Mas, S. K. Solanki, and M. Stix, \aa\ {\bf 314}, 503 (1996).

\bibitem[105]{isik}  
  E. Isik, D. Schmitt, and M. Sch\"ussler, \aa\ {\bf 528}, 135 (2011).
  
\bibitem[106]{kit16}
  L. L. Kitchatinov, and S. V. Olemskoy, \mnras\ {\bf 459}, 4353 (2016).
  
\bibitem[107]{mest}
  L. Mestel, {\em Stellar Magnetism}, Clarendon Press, Oxford (1999).
  
\bibitem[108]{don09}
  J.-F. Donati, and J. D. Landstreet, J. D., Ann. Rev. Astron. Astrophys. {\bf 47}, 333 (2009).
  
\bibitem[109]{cho02}
  A. R. Choudhuri and S. Konar, \mnras\ {\bf 332}, 933 (2002).

\bibitem[110]{cho04}
  S. Konar and A. R. Choudhuri, \mnras\ {\bf 348}, 661 (2004).
  
\bibitem[111]{see}
  V. See, et al., \mnras\ {\bf 462}, 4442 (2016).
  
\bibitem[112]{cho16}  
  A. R. Choudhuri, and G. Hazra, Adv. Space Res. {\bf 58}, 1560 (2016).
  
\bibitem[113]{yea}
  A. R. Yeates, and A. Mu\~noz-Jaramillo, \mnras\ {\bf 436}, 3366 (2013).
  
\bibitem[114]{mie}
  M. S. Miesch, and M. Dikpati, \apj\ {\bf 785}, L8 (2014).
  
\bibitem[115]{haz16}
  G. Hazra, A. R. Choudhuri, and M. S. Miesch, \apj\ (submitted, arXiv:1610.02726) (2017).
  
\bibitem[116]{brown}  
  B. P. Brown, M. K. Browning, A. S. Brun, M. S. Miesch, and J. Toomre, \apj\ {\bf 711}, 424 (2010).
  
\bibitem[117]{ghi}
  M. Ghizaru, P. Charbonneau, and P. K. Smolarkiewicz, \apj\ {\bf 715}, L133 (2010).
  
\end{thebibliography}
\end{document}